\PassOptionsToPackage{unicode}{hyperref}
\PassOptionsToPackage{hyphens}{url}
\PassOptionsToPackage{dvipsnames,svgnames,x11names}{xcolor}
\documentclass[
  12pt]{article}

\usepackage{amsmath,amssymb}
\usepackage{amsthm}
\usepackage{iftex}
\ifPDFTeX
  \usepackage[T1]{fontenc}
  \usepackage[utf8]{inputenc}
  \usepackage{textcomp} % provide euro and other symbols
\else % if luatex or xetex
  \usepackage{unicode-math}

  \defaultfontfeatures{Scale=MatchLowercase}
  \defaultfontfeatures[\rmfamily]{Ligatures=TeX,Scale=1}
\fi
\usepackage{lmodern}
\ifPDFTeX\else  
\fi
\IfFileExists{upquote.sty}{\usepackage{upquote}}{}
\IfFileExists{microtype.sty}{% use microtype if available
  \usepackage[]{microtype}
  \UseMicrotypeSet[protrusion]{basicmath} % disable protrusion for tt fonts
}{}
\makeatletter
\@ifundefined{KOMAClassName}{% if non-KOMA class
  \IfFileExists{parskip.sty}{%
    \usepackage{parskip}
  }{% else
    \setlength{\parindent}{0pt}
    \setlength{\parskip}{6pt plus 2pt minus 1pt}}
}{% if KOMA class
  \KOMAoptions{parskip=half}}
\makeatother
\usepackage{xcolor}
\makeatletter
\ifx\paragraph\undefined\else
  \let\oldparagraph\paragraph
  \renewcommand{\paragraph}{
    \@ifstar
      \xxxParagraphStar
      \xxxParagraphNoStar
  }
  \newcommand{\xxxParagraphStar}[1]{\oldparagraph*{#1}\mbox{}}
  \newcommand{\xxxParagraphNoStar}[1]{\oldparagraph{#1}\mbox{}}
\fi
\ifx\subparagraph\undefined\else
  \let\oldsubparagraph\subparagraph
  \renewcommand{\subparagraph}{
    \@ifstar
      \xxxSubParagraphStar
      \xxxSubParagraphNoStar
  }
  \newcommand{\xxxSubParagraphStar}[1]{\oldsubparagraph*{#1}\mbox{}}
  \newcommand{\xxxSubParagraphNoStar}[1]{\oldsubparagraph{#1}\mbox{}}
\fi
\makeatother

\usepackage{longtable,booktabs,array}
\usepackage{calc} % for calculating minipage widths
\usepackage{etoolbox}
\makeatletter
\patchcmd\longtable{\par}{\if@noskipsec\mbox{}\fi\par}{}{}
\makeatother
\IfFileExists{footnotehyper.sty}{\usepackage{footnotehyper}}{\usepackage{footnote}}
\makesavenoteenv{longtable}
\usepackage{graphicx}
\makeatletter
\def\maxwidth{\ifdim\Gin@nat@width>\linewidth\linewidth\else\Gin@nat@width\fi}
\def\maxheight{\ifdim\Gin@nat@height>\textheight\textheight\else\Gin@nat@height\fi}
\makeatother
\setkeys{Gin}{width=\maxwidth,height=\maxheight,keepaspectratio}
\makeatletter
\def\fps@figure{htbp}
\makeatother

\makeatletter
\@ifpackageloaded{caption}{}{\usepackage{caption}}
\AtBeginDocument{%
\ifdefined\contentsname
  \renewcommand*\contentsname{Table of contents}
\else
  \newcommand\contentsname{Table of contents}
\fi
\ifdefined\listfigurename
  \renewcommand*\listfigurename{List of Figures}
\else
  \newcommand\listfigurename{List of Figures}
\fi
\ifdefined\listtablename
  \renewcommand*\listtablename{List of Tables}
\else
  \newcommand\listtablename{List of Tables}
\fi
\ifdefined\figurename
  \renewcommand*\figurename{Figure}
\else
  \newcommand\figurename{Figure}
\fi
\ifdefined\tablename
  \renewcommand*\tablename{Table}
\else
  \newcommand\tablename{Table}
\fi
}
\@ifpackageloaded{float}{}{\usepackage{float}}
\floatstyle{ruled}
\@ifundefined{c@chapter}{\newfloat{codelisting}{h}{lop}}{\newfloat{codelisting}{h}{lop}[chapter]}
\floatname{codelisting}{Listing}

\makeatother
\makeatletter
\@ifpackageloaded{caption}{}{\usepackage{caption}}
\@ifpackageloaded{subcaption}{}{\usepackage{subcaption}}
\makeatother

\ifLuaTeX
  \usepackage{selnolig}  % disable illegal ligatures
\fi
\usepackage[]{natbib}
\usepackage{bookmark}

\IfFileExists{xurl.sty}{\usepackage{xurl}}{} % add URL line breaks if available
\hypersetup{
  pdftitle={Randomization Inference for Local Projections with Persistent Instruments in Short Panels},
  pdfauthor={Pedro Picchetti},
  pdfkeywords={Local projections, Instrumental variables, Randomization inference, Panel data, Persistent instruments, Nuisance parameters},
  colorlinks=true,
  linkcolor={blue},
  filecolor={Maroon},
  citecolor={Blue},
  urlcolor={Blue},
  pdfcreator={LaTeX via pandoc}}

\usepackage{algorithm}
\usepackage{algpseudocode}

\newcommand{\Var}{\mathbb{V}}
\newcommand{\Cov}{\mathbb{C}\mathrm{ov}}
\newcommand{\E}{\mathbb{E}}
\newcommand{\Prob}{\mathbb{P}}
\newcommand{\R}{\mathcal{R}}
\newcommand{\G}{\mathcal{G}}
\newcommand{\ind}{\mathbf{1}}

\newcommand{\anon}{1}
\newtheorem{proposition}{Proposition}

\newtheorem{theorem}{Theorem}
\newtheorem{assumption}{Assumption}
\theoremstyle{definition}
\newtheorem{definition}{Definition}
\theoremstyle{remark}
\newtheorem{remark}{Remark}

\begin{document}

\def\spacingset#1{\renewcommand{\baselinestretch}%
{#1}\small\normalsize} \spacingset{1}

%%%%%%%%%%%%%%%%%%%%%%%%%%%%%%%%%%%%%%%%%%%%%%%%%%%%%%%%%%%%%%%%%%%%%%%%%%%%%%

\if1\anon
{
  \title{\bf Randomization Inference for Dynamic Causal Effects in Short Panels with Autoregressive Instruments}
  \author{Pedro Picchetti\thanks{
    pedro.picchetti@uc.cl.}\hspace{.2cm}\\
    Institute of Economics, PUC-Chile\\
    }
  \maketitle
} \fi

\if0\anon
{
  \bigskip
  \bigskip
  \bigskip
  \begin{center}
    {\LARGE\bf Randomization Inference for Dynamic Causal Effects in Short Panels with Autoregressive Instruments}
\end{center}
  \medskip
} \fi

\bigskip
\begin{abstract}
I study randomization inference for dynamic causal effects in panels with many units and few periods when treatment is endogenous and an instrument is as good as randomly assigned only through the innovations of a first-order autoregressive assignment mechanism with unknown persistence. The persistence is a nuisance parameter in two senses: it determines which weighted average of heterogeneous causal effects the instrumented local projection estimand measures, and it governs the randomization distribution of any statistic built from the instrument. I propose tests that never estimate it. For each candidate value the innovations are recovered, rerandomized under a group that leaves their joint law invariant, and a randomization p-value is formed; the reported p-value is the supremum over candidate values or over an exact confidence set. Under a sharp null on the path of lag-specific effects the tests have level at most $\alpha$ for every persistence in finite samples, including the unit root, and the supremum is a maximum over finitely many points.  Under a weak null on the heterogeneity-weighted estimand a studentized version is uniformly asymptotically valid. 
\end{abstract}

\noindent%
\vfill

\newpage
\spacingset{1.8} % DON'T change the spacing!

\section{Introduction}\label{sec-intro}

Local projections \citep{jorda2005} are a standard tool for the estimation of dynamic causal effects using time series and panel data. When treatments are endogenous, an external shock provides exogenous variations that allows the identification of causal effects. In such applications, the instrument is rarely a white noise. Cloud cover, weather, commodity prices, narrative policy shocks, or
exposure to an aggregate shock all display persistence over time.

This paper focuses on short panel settings where persistent instruments have an autoregressive format, $Z_{i,t}=\rho_{0} Z_{i,t-1}+\varepsilon_{i,t}$, and only the innovations $\varepsilon_{i,t}$ are as good as randomly assigned. Persistence creates two
distinct problems. First, it affects the causal interpretation of the instrumented local projections estimand. with a serially correlated instrument, the horizon-$h$ coefficient loads on the effects of past and future
treatments as well as on the horizon-$h$ effect.

The second problem is associated to the inference. a design-based test rerandomizes the assignment, and the assignment that is rerandomizable is the innovation array, which is only recovered once $\rho$ is known. Plugging in an estimate is unattractive in exactly the setting we consider. With few periods and unit heterogeneity in the instrument process, the within-unit estimate of $\rho$ is biased by a term of order $(1+\rho)/T$ \citep{nickell1981}, and fixed-effects local projections inherit a bias of the same kind \citep{herbstjohannsen2024,meishengshi2026}.

In this paper, I study how to test hypotheses about causal effects, and how to construct confidence sets, with guarantees that do not depend on the unknown persistence $\rho_{0}$.

I propose to treat $\rho_{0}$ as a  nuisance parameter of the assignment mechanism, and to handle it in the manner of \citet{dufour2006} and \citet{bergerboos1994}. For each candidate value $\rho$, recover the innovations $\varepsilon_{i,t}(\rho)=Z_{i,t}-\rho Z_{i,t-1}$, apply a group of transformations under which the true innovation array is invariant in law, such as permutations of innovations across units within each period, and compute a randomization $p$-value $p(\rho)$. 

At the true value the $p$-value is exact. Since the true value is unknown we report $\sup_\rho p(\rho)$, or the supremum over an exact confidence set for $\rho_{0}$ plus the confidence set's error probability. Two features make this more than a textbook application of maximized Monte Carlo tests. First, the test statistic is a polynomial of low degree in $\rho$, so $p(\cdot)$ is a step function with finitely many breakpoints that can be located exactly; the supremum is a maximum over a finite set, and no grid approximation is needed. Second, the fixed-$T$, large-$N$ structure means that the unit root is just another point in the parameter space: the limit theory is a cross-sectional central limit theorem, uniform in $\rho$ over any compact set, and the discontinuities that plague autoregressive inference as $T\to\infty$ never arise.

Under a \emph{sharp} null that fixes a common impulse-response path and so imputes all counterfactual outcomes, Theorem~\ref{thm:exact} shows that the maximized $p$-value has level at most $\alpha$ for every $\rho$ in finite samples, for any $N$, $T$ and number of Monte Carlo draws, with no restriction on the first stage and no moment assumptions. Under a \emph{weak} null that fixes only the heterogeneity-weighted estimand, for which no randomization test is exact \citep{chungromano2013,wuding2021}, Theorem~\ref{thm:asym} shows that the studentized version of the procedure is asymptotically valid as $N\to\infty$ with $T$ fixed, uniformly over $\rho$. Proposition~\ref{prop:power} gives consistency of the maximized test and identifies the exceptional alternatives against which a single-horizon test has no power; Proposition~\ref{prop:fe} treats unit effects in the outcome and in the assignment mechanism through within-unit stratified permutation schemes. Proposition~\ref{prop:strat} extends the assignment mechanism to stratified permutation schemes, and Proposition \ref{prop:cluster} extend the results to clustered designs.

\paragraph*{Related work.} I build on the design-based analysis of dynamic causal effects in the potential outcomes tradition of \citet{neyman1923} and \citet{fisher1935}: \citet{bojinovshephard2019} develop exact randomization tests for time-series experiments with known adaptive assignment probabilities, \citet{bojinov} extend them to panel experiments, and \citet{rambachanshephard} give nonparametric causal meaning to local projection estimands, including the instrumented version, under random assignment of the instrument. The design-based view of quasi-experiments with unknown assignment probabilities is developed by \citet{jonashesh}. This paper's departure from this setting is that the assignment mechanism is known only up to a parameter that cannot be estimated reliably. The closest precedent is \citet{shaikhtoulis2021}, who study Fisher tests under a Cox adoption model: finite-sample validity holds with known parameters and asymptotic validity with plug-in estimates.

I obtain finite-sample validity for every value of the unknown parameter, following the maximized Monte Carlo approach of \citet{dufour2006}, the exact autoregressive inference of \citet{dufourkiviet1998}, and the confidence-set approach of \citet{bergerboos1994}. The distinction between exact tests of sharp nulls and asymptotically valid studentized tests of weak nulls follows \citet{chungromano2013}, \citet{diciccioromano2017}, \citet{canayromanoshaikh2017} and \citet{wuding2021}.
The large-$N$ statements are of the uniform kind advocated by \citet{andrewsguggenberger2010}, and the supremum over $\rho$ is in the spirit of the grid bootstrap of \citet{hansen1999}.

The estimand results relate to \citet{stockwatson2018}, \citet{plagborgwolf2021} and \citet{allozagonzalosanz2025}, who study persistent instruments and shocks in time series with homogeneous effects, and to the weakly causal interpretation of two-stage least squares under heterogeneity of \citet{blandhol}. The filtered instrument is the policy innovation of \citet{angristkuersteiner2011} and \citet{angristjordakuersteiner2018} and the recentered instrument of \citet{borusyakhull2023}, who also rerandomize from a known assignment process.

Section~\ref{sec-setup} sets up the framework and the assignment mechanism. Section~\ref{sec-estimand} provides the causal decomposition of the instrumented local projection estimand and its filtered counterpart. Section~\ref{sec-inference} develops the tests. Sections~\ref{sec-mc} and~\ref{sec-app} contain simulations and the application to satellite-based environmental enforcement in the Brazilian Amazon, where cloud cover is the persistent instrument. Section~\ref{sec-conclusion} concludes. All proofs are in the Supplementary Material.

\section{Framework}\label{sec-setup}

\subsection{Potential outcomes}

Consider a balanced panel in which $N$ units are observed over $T$ periods of time. For each unit $i\in[N]$ and time period $t\in[T]$, we observe an instrumental variable $Z_{i,t}$, a treatment variable $X_{i,t}$ and a real-valued scalar outcome $Y_{i,t}$.

\begin{assumption}[No Spillovers and No Anticipation]\label{ass:sutva}
For all $i\in[N]$ and $t\in[T]$, $Y_{i,t}(x_{1:N,1:T},z_{1:N,1:T})=Y_{i,t}(x_{i,1:t},z_{i,1:t})$ and $X_{i,t}(z_{1:N,1:T})=X_{i,t}(z_{i,1:t})$.
\end{assumption}

Assumption 1 imposes that the potential outcome of unit $i$ in period $t$ depends only on the treatment and instrument paths of unit $i$ until period $t$, ruling out the possibility of spillover of both treatments and instruments across units, as well as future treatments affecting past potential outcomes. It also imposes that the potential treatment from unit $i$ in period $t$ depends only on the instrument paths of unit $i$ until period $t$.

One of the fundamental assumptions of IV settings is the exclusion restriction. Its dynamic version is stated below, alongside an additional exclusion restriction for the first stage.

\begin{assumption}[Exclusion Restrictions]\label{ass:excl}
For all $i\in[N]$ and $t\in[T]$, (i) $Y_{i,t}(x_{i,1:t},z_{i,1:t})=Y_{i,t}(x_{i,1:t})$ and (ii) $X_{i,t}(z_{i,1:t})=X_{i,t}(z_{i,t})$.
\end{assumption}

The first part of Assumption 2 is the standard exclusion restriction for dynamic IV settings. It states that the path of instruments does not affect potential outcomes directly. Instruments only affect potential outcomes to the extent that they affect treatment choices. Mapping it to the application, it implies that the history of cloud coverage over a municipality affects illegal deforestation in that municipality only through its effect on law enforcement capabilities.

The second part of Assumption 2 imposes that potential treatments in period $t$ depend only on the instruments in period $t$. Although it might seem strong in a dynamic setting, it is justifiable in many experimental and observational settings. In AGR, the capacity of the satellite to detect illegal deforestation at a given period depends on the cloud coverage below the satellite at that period. Previous coverage affects the previous law enforcement capacity but does not affect the current capacity to detect deforestation.

Most applications of IV methods with panel data use linear estimators, such as the pooled 2SLS, or 2SLS with fixed effects. In the continuous treatment/instrument setting, I assume that potential outcomes are a linear function of the treatment path, and potential treatments are a linear function of the contemporary instrument.

\begin{assumption}[Linear potential quantities]\label{ass:linear}
For all $i\in[N]$ and $t\in[T]$,
$Y_{i,t}(x_{i,1:t})=Y_{i,t}(\mathbf{0})+\sum_{p=0}^{t-1}\beta_{i,t,p}\,x_{i,t-p}$ and $X_{i,t}(z_{i,t})=X_{i,t}(0)+\pi_{i,t}\,z_{i,t}$.
\end{assumption}

The coefficients $\beta_{i,t,p}$ capture the causal effect of an increase of treatment from $p$ for unit $i$ in period $t$. For instance, $\beta_{i,t,0}$ captures the causal effect of the contemporary treatment for unit $i$ in period $t$, and $\beta_{i,t,1}$ captures the causal effect of the treatment in $t-1$ for unit $i$ in period $t$.

The fundamental behavioral assumption in IV settings is the monotonicity assumption, which is provided below.

\begin{assumption}[Monotonicity]\label{ass:mono}
For all $i\in[N]$ and $t\in[T]$, $\pi_{i,t}\geq 0$.
\end{assumption}

\subsection{Assignment mechanism}

Let $\mathcal{F}_{t-1}$ denote the sigma-field generated by all potential quantities, the initial conditions $Z_{1:N,0}$, and $Z_{1:N,1:t-1}$. We assume that the instrument follows an AR(1) process.

\begin{assumption}[AR(1) assignment mechanism with exogenous innovations]\label{ass:ar1}
There exist $\rho_{0}\in\R$, with $\R\subset\mathbb{R}$ a known compact set, and an array $\varepsilon=(\varepsilon_{i,t})_{i\in[N],t\in[T]}$ such that for all $i,t$,
\begin{equation*}
Z_{i,t}=\rho_{0} Z_{i,t-1}+\varepsilon_{i,t},\qquad \E\left[\varepsilon_{i,t}\mid\mathcal{F}_{t-1}\right]=0,\qquad \Var\left(\varepsilon_{i,t}\mid\mathcal{F}_{t-1}\right)=\sigma^{2}_{i,t}\in(0,\infty).
\end{equation*}
\end{assumption}

The leading case is $\R=[-1,1]$. However, because $T$ is fixed, neither $|\rho_{0}|<1$ nor stationarity is required. Assumption~\ref{ass:ar1} is a design-based sequential ignorability condition \citep{ROBINS19861393,murphy}: the innovation is mean independent of potential quantities and of the past of the instrument, while the initial conditions are held fixed by conditioning. Innovation variances may differ across units and periods, and no independence across units is imposed at this stage. Unit-specific means and seasonal components in the assignment mechanism are treated in Sections~\ref{sec-fe} and~\ref{sec-general}.

\begin{definition}[Filtered instrument at a candidate persistence]\label{def:filtered}
For $\rho\in\R$, $\varepsilon_{i,t}(\rho):=Z_{i,t}-\rho Z_{i,t-1}$. Under Assumption~\ref{ass:ar1}, $\varepsilon_{i,t}(\rho_{0})=\varepsilon_{i,t}$ and $\varepsilon_{i,t}(\rho)=\varepsilon_{i,t}+(\rho_{0}-\rho)Z_{i,t-1}$.
\end{definition}

For each $\rho$ the map $(Z_{i,0},\varepsilon_{i,1:T})\mapsto Z_{i,0:T}$ is a bijection, so a rerandomized innovation array $\varepsilon^{*}$ corresponds to a unique rerandomized instrument path $Z^{*}_{i,t}(\rho)=\rho Z^{*}_{i,t-1}(\rho)+\varepsilon^{*}_{i,t}$, with $Z^{*}_{i,0}(\rho)=Z_{i,0}$.

\section{Estimands}\label{sec-estimand}

In this section, I derive the causal decomposition of local projection estimands. Expectations are taken over the innovations, conditional on potential quantities and initial conditions; sums over $t$ at horizon $h$ run over $\mathcal{T}_h:=\{1,\dots,T-h\}$.

The instrumented local projection (LP-IV) estimand at horizon $h$ and its filtered counterpart are
\begin{equation}\label{eq:theta}
\beta^{\mathrm{LP}}_{h}:=\frac{\sum_{t\in\mathcal{T}_h}\sum_{i}\E\left[Z_{i,t}Y_{i,t+h}\right]}{\sum_{t\in\mathcal{T}_h}\sum_{i}\E\left[Z_{i,t}X_{i,t}\right]},
\qquad
\theta_{h}:=\frac{\sum_{t\in\mathcal{T}_h}\sum_{i}\E\left[\varepsilon_{i,t}Y_{i,t+h}\right]}{\sum_{t\in\mathcal{T}_h}\sum_{i}\E\left[\varepsilon_{i,t}X_{i,t}\right]},
\end{equation}
the population analogues of a just-identified instrumental-variables regression of $Y_{i,t+h}$ on $X_{i,t}$ pooled over units and periods. At $h=0$ the first is the static pooled two-stage least squares estimand. The following result does not use the autoregressive structure; it requires only mean-zero exogenous instruments. Under the autoregressive structure, instruments have mean zero if \ref{ass:ar1} holds and $\mathbb{E}\left[Z_{i,0}\right]=0$ for all $i\in[N]$.

\begin{proposition}[LP-IV under a serially correlated instrument]\label{prop:unfiltered}
Suppose Assumptions~\ref{ass:sutva}--\ref{ass:mono} hold, $\E[Z_{i,t}]=0$ for all $i,t$, and $\sum_{t\in\mathcal{T}_h}\sum_i\Var(Z_{i,t})\pi_{i,t}\neq0$. Then
\begin{equation}\label{eq:unfiltered}
\begin{gathered}
\beta^{\mathrm{LP}}_{h}=\sum_{t\in\mathcal{T}_h}\sum_{i}\lambda_{i,t}\,\beta_{i,t+h,h}
+\sum_{t\in\mathcal{T}_h}\sum_{i}\sum_{p\neq h}\lambda_{i,t,p}\,\beta_{i,t+h,p},\\
\lambda_{i,t}=\frac{\Var(Z_{i,t})\pi_{i,t}}{\sum_{s\in\mathcal{T}_h}\sum_{j}\Var(Z_{j,s})\pi_{j,s}},
\qquad
\lambda_{i,t,p}=\frac{\Cov(Z_{i,t},Z_{i,t+h-p})\pi_{i,t+h-p}}{\sum_{s\in\mathcal{T}_h}\sum_{j}\Var(Z_{j,s})\pi_{j,s}},
\end{gathered}
\end{equation}
where $p$ ranges over $\{0,\dots,t+h-1\}\setminus\{h\}$. 
\end{proposition}

The LP-IV estimand can be decomposed into a convex weighted average of causal effects $\beta_{i,t+h,h}$ and a linear combination of other dynamic causal effects in which the coefficients depend on the serial correlation of the instrument. In order for the estimand to hold a \textit{weakly causal} interpretation, it must be the case the for all $i\in[N]$ $t\in [T]$, the instrument $Z_{i,t}$ is positively correlated with its lags and leads.

The serial correlation of the instrument is estimable, so a researcher can check whether the weights $\lambda_{i,t,p}$ are negative; but even when they are all positive, as under an autoregression with $\rho_{0}>0$, the horizon-$h$ effect cannot be separated from the effects at other horizons. Replacing the instrument by its innovation removes the past-treatment terms but not the future-treatment terms, as shown in Proposition 2.

\begin{proposition}[Filtered LP-IV under the AR(1) mechanism]\label{prop:filtered}
Suppose Assumptions~\ref{ass:sutva}--\ref{ass:linear} and~\ref{ass:ar1} hold and $\sum_{t\in\mathcal{T}_h}\sum_i\sigma^{2}_{i,t}\pi_{i,t}\neq0$. Then
\begin{equation}\label{eq:decomp}
\theta_{h}=\sum_{t\in\mathcal{T}_h}\sum_{i}w_{i,t}\,\beta_{i,t+h,h}
+\sum_{t\in\mathcal{T}_h}\sum_{i}\sum_{p=0}^{h-1}\omega_{i,t,p}\,\beta_{i,t+h,p},
\quad
w_{i,t}=\frac{\sigma^{2}_{i,t}\pi_{i,t}}{\sum_{t\in\mathcal{T}_h}\sum_i\sigma^{2}_{i,t}\pi_{i,t}},
\quad
\omega_{i,t,p}=\frac{\rho^{\,h-p}\sigma^{2}_{i,t}\pi_{i,t+h-p}}{\sum_{t\in\mathcal{T}_h}\sum_i\sigma^{2}_{i,t}\pi_{i,t}}.
\end{equation}
\end{proposition}

The innovation $\varepsilon_{i,t}$ is uncorrelated with lagged instruments, which removes the past-treatment terms of~\eqref{eq:unfiltered}, but it is propagated forward by the autoregression, $\Cov(\varepsilon_{i,t},Z_{i,t+j})=\rho_{0}^{j}\sigma^{2}_{i,t}$, so it shifts the treatments in periods $t+1,\dots,t+h$, to which the horizon-$h$ outcome responds through $\beta_{i,t+h,p}$, $p<h$. For $h\geq1$ and $\rho_{0}\neq0$, $\theta_h$ is therefore not a convex combination of horizon-$h$ effects. The natural reading of $\theta_h$, valid under arbitrary heterogeneity, is as a relative impulse response: the ratio of the average response of $Y_{t+h}$ to the instrument innovation to the average impact response of $X_{it}$ \citep{stockwatson2018,plagborgwolf2021}. 

\begin{remark}[Unit and period effects in the estimand]\label{rem:fe-estimand}
Applied work typically includes unit and period effects, replacing the instrument by its within-unit transformed version $\breve Z_{i,t}=Z_{i,t}-|\mathcal{T}_h|^{-1}\sum_{s\in\mathcal{T}_h}Z_{i,s}$. The decomposition of the resulting estimand, Proposition S.1 in the Supplementary Material, contains two additional terms, $\sum_{t,i}\E[\breve Z_{i,t}]\,Y_{i,t+h}(\mathbf{0})$ and $\sum_{t,i,p}\E[\breve Z_{i,t}]\,X_{i,t+h-p}(0)\beta_{i,t+h,p}$, and modified weights involving $\Cov(\breve Z_{i,t},Z_{i,t+h-p})$. The additional terms vanish when $\E[Z_{i,t}]$ is constant over $t$ within each unit, or when $Y_{i,t}(\mathbf{0})$ and $X_{i,t}(0)$ are additive in unit and period effects; under Assumption~\ref{ass:ar1} with initial conditions held fixed, however, $\E[Z_{i,t}]=\rho^{t}Z_{i,0}$ varies with $t$, so with $T$ fixed the within transformation of the instrument is not innocuous. This is a design-based form of the incidental-parameter problem. The randomization tests of Section~\ref{sec-inference} do not transform the instrument within units, and under the sharp null they accommodate arbitrary $Y_{i,t}(\mathbf{0})$, so unit effects in outcomes cost nothing.
\end{remark}

Proposition 2 shows that the impulse response function is recovered from the filtered coefficients only through $\rho_{0}$. The parameter that must be known to construct the filtered instrument is also needed to interpret what it identifies. That provides the motivation for inference that holds for every $\rho$ at once.

\section{Randomization Inference without estimating
\texorpdfstring{$\rho_{0}$}{rho_0}}
\label{sec-inference}
\subsection{Sharp and weak null hypotheses}

Fix a maximal horizon $H\leq T-1$. Randomization tests require the null to determine the outcomes that would have been observed under any rerandomized assignment.

\begin{definition}[Sharp null on the impulse-response path]
For $b=(b_0,\dots,b_H)\in\mathbb{R}^{H+1}$, $\mathbb{H}^{s}_0(b)$ states that for all $i\in[N]$, $t\in[T]$ and all treatment paths,
$Y_{i,t}(x_{i,1:t})=Y_{i,t}(\mathbf{0})+\sum_{p=0}^{\min\{t-1,H\}}b_p\,x_{i,t-p}$.
\end{definition}

$\mathbb{H}^{s}_0(b)$ imposes a common linear impulse-response path with memory $H$ and leaves $Y_{i,t}(\mathbf{0})$ and the first stage unrestricted. Typically, researchers are interested on the Fisher null of no effects, $b=\textbf{0}$.

Under $\mathbb{H}_{0}^{s}(b)$, the adjusted outcomes
\begin{equation}\label{eq:W}
W_{i,t}(b):=Y_{i,t}-\sum_{p=0}^{\min\{t-1,H\}}b_p\,X_{i,t-p}
\end{equation}
equal $Y_{i,t}(\mathbf{0})$ and are nonstochastic. For the fisher null, $W_{i,t}(b)=Y_{i,t}=Y_{i,t}(\mathbf{0})$. Inverting the test over $b$ gives a confidence set for the path; sets for a single $b_h$ follow by projection. 

\begin{definition}[Weak null on the filtered estimand]
For $h\leq H$ and $\theta\in\mathbb{R}$, $\mathbb{H}^{w}_0(h,\theta)$ states that $\theta_h=\theta$, with $\theta_h$ defined in~\eqref{eq:theta} at the true $\rho_{0}$.
\end{definition}

$\mathbb{H}^{w}_0(h,\theta)$ allows arbitrary heterogeneity in $\beta_{i,t,p}$ and $\pi_{i,t}$ and does not determine counterfactuals. We use the adjusted outcomes $V_{i,t+h}(\theta):=Y_{i,t+h}-\theta X_{i,t}$, $t\in\mathcal{T}_h$, for which $\sum_{t\in\mathcal{T}_h}\sum_i\Cov(\varepsilon_{i,t},V_{i,t+h}(\theta))=0$ under $\mathbb{H}^{w}_0(h,\theta)$.

\subsection{Randomization hypothesis, statistics and $p$-values}\label{sec-pvalues}

Let $\G$ be a finite group of transformations of $N\times T$ arrays, $g\varepsilon$ the transformed array.

\begin{assumption}[Randomization hypothesis]\label{ass:invariance}
Conditional on potential quantities and initial conditions, $g\varepsilon\overset{d}{=}\varepsilon$ for every $g\in\G$.
\end{assumption}

The \emph{cross-sectional group} $\G_{\mathrm{cs}}$ consists of $g=(\sigma_1,\dots,\sigma_T)$, $\sigma_t$ a permutation of $[N]$, acting by $(g\varepsilon)_{i,t}=\varepsilon_{\sigma_t(i),t}$. Assumption~\ref{ass:invariance} then requires innovations to be exchangeable across units within each period, as when they are independent across units with a period-specific distribution, possibly with common period shocks. The \emph{within-unit group} $\G_{\mathrm{wu}}$ consists of $g=(\tau_1,\dots,\tau_N)$, $\tau_i$ a permutation of $[T]$, acting by $(g\varepsilon)_{i,t}=\varepsilon_{i,\tau_i(t)}$, and requires exchangeability over time within each unit, allowing unit-specific innovation laws. Stratified versions, in which permutations act within blocks of a partition of $[N]\times[T]$, are covered in Section~\ref{sec-fe}. We present the main results for $\G_{\mathrm{cs}}$.

For an array $a$, $\bar a_t=N^{-1}\sum_ia_{i,t}$ and $\tilde a_{i,t}=a_{i,t}-\bar a_t$ denote within-period demeaning. For $r\in\R$ and an array $U$ of adjusted outcomes, equal to $W(b)$ or $V(\theta)$, define
\begin{equation}\label{eq:stat}
\begin{gathered}
T_h(\rho;U)=\sum_{t\in\mathcal{T}_h}\sum_{i=1}^{N}\tilde\varepsilon_{i,t}(\rho)\,\tilde U_{i,t+h},
\qquad
\hat s^{2}_h(\rho;U)=\sum_{i=1}^{N}\Big(\sum_{t\in\mathcal{T}_h}\tilde\varepsilon_{i,t}(\rho)\,\tilde U_{i,t+h}\Big)^{2},\\
S_h(\rho;U)=\frac{|T_h(\rho;U)|}{\hat s_h(\rho;U)},
\qquad
S(\rho;U)=\max_{0\leq h\leq H}S_h(\rho;U).
\end{gathered}
\end{equation}
For the sharp null any measurable function of $(\varepsilon(\rho),W(b))$ is admissible; we use $S(\rho;W(b))$. For the weak null we use $S_h(\rho;V(\theta))$. Transformed statistics replace $\varepsilon(\rho)$ by $g\varepsilon(\rho)$; since $\G_{\mathrm{cs}}$ permutes within periods, $\tilde\varepsilon(\rho)$ is permuted in the same way. Because the first stage is unknown, a Wald statistic based on the IV estimator cannot be rerandomized (the treatment under a rerandomized path, $X_{i,t}(0)+\pi_{i,t}Z^{*}_{i,t}(\rho)$, depends on $\pi_{i,t}$); the reduced-form statistic avoids this and is robust to weak first stages \citep{andersonrubin1949}.

The procedure is as follows: draw $g_1,\dots,g_M$ independently and uniformly from $\G$ once, and use the same draws for every $\rho$. The randomization $p$-value at $\rho$, the \emph{maximized Monte Carlo} (MMC) $p$-value, and the \emph{confidence-set} $p$-value are, respectively,
\begin{equation}\label{eq:pvals}
\begin{gathered}
p_M(\rho)=\frac{1+\sum_{m=1}^{M}\ind\{S(g_m\varepsilon(\rho);U)\geq S(\varepsilon(\rho);U)\}}{M+1},\\
p^{\sup}_{M}=\sup_{\rho\in\R}p_M(\rho),
\qquad
p^{\mathrm{BB}}_{M,\gamma}=\sup_{\rho\in C_{1-\gamma}}p_M(\rho)+\gamma,
\end{gathered}
\end{equation}
with $\sup_\emptyset=0$, $\gamma\in(0,\alpha)$, and $C_{1-\gamma}=\{\rho\in\R:p^{\rho_{0}}_M(\rho)>\gamma\}$ an exact randomization confidence set for $\rho_{0}$ obtained by inverting a test of the hypothesis that $\varepsilon(\rho)$ is $\G$-invariant. For the latter we use the orthogonality statistic
\begin{equation}\label{eq:aux}
B(\rho)=\Big|\sum_{t=1}^{T}\sum_{i}\tilde\varepsilon_{i,t}(\rho)\,\tilde Z_{i,t-1}\Big|,
\end{equation}
whose transformed version $B(g\varepsilon(\rho))$ is computed with the lagged instrument rebuilt from the transformed innovations, $Z^{*}_{i,t-1}(\rho)$, so that it is a function of $(g\varepsilon(\rho),Z_{1:N,0})$ alone; $p^{\rho_{0}}_M(r)$ is its $p$-value as in~\eqref{eq:pvals}. At $\rho=\rho_{0}$ the innovations are orthogonal to the lagged instrument, while $\E[\varepsilon_{i,t}(\rho)Z_{i,t-1}]=(\rho_{0}-\rho)\E[Z^{2}_{i,t-1}]$ is nonzero for every $\rho\neq\rho_{0}$. The test rejects when the chosen $p$-value is at most $\alpha$; Algorithm~\ref{alg:main} summarizes the computation.

\begin{algorithm}[H]
\spacingset{1.15}
\caption{$\rho$-free randomization $p$-value}\label{alg:main}
\begin{algorithmic}[1]
\Require Panel $\{Z_{i,0:T},X_{i,1:T},Y_{i,1:T}\}_{i=1}^{N}$; null ($b$ for $H^{s}_0$, or $(h,\theta)$ for $H^{w}_0$); group $\G$; $M$; $\R$; $\alpha$; optionally $\gamma\in(0,\alpha)$.
\State Draw $g_1,\dots,g_M$ i.i.d. uniform on $\G$ and store them.
\State Form adjusted outcomes $U=W(b)$ by~\eqref{eq:W} or $U=V(\theta)$.
\State Candidate set $\R^{\dagger}\subset\R$: the endpoints of $\R$ (or a fine grid).
\For{each $\rho\in\R^{\dagger}$}
  \State $\varepsilon_{i,t}(\rho)\gets Z_{i,t}-rZ_{i,t-1}$, demeaned within periods.
  \State Compute $S(\varepsilon(\rho);U)$ and $S(g_m\varepsilon(\rho);U)$, $m\leq M$, by~\eqref{eq:stat}; set $p_M(\rho)$ by~\eqref{eq:pvals}.
  \If{$\gamma$ is supplied} rebuild $Z^{*}(\rho)$ from $g_m\varepsilon(\rho)$, compute $B(\rho)$, $B(g_m\varepsilon(\rho))$ and $p^{\rho_{0}}_M(\rho)$. \EndIf
\EndFor
\State $p^{\sup}_M\gets\max_{\rho\in\R^{\dagger}}p_M(\rho)$; if $\gamma$ is supplied, $C_{1-\gamma}\gets\{\rho\in\R^{\dagger}:p^{\rho_{0}}_M(\rho)>\gamma\}$ and $p^{\mathrm{BB}}_{M,\gamma}\gets\max_{\rho\in C_{1-\gamma}}p_M(\rho)+\gamma$.
\State Reject if the chosen $p$-value is at most $\alpha$; confidence sets collect the non-rejected $b$ (or $\theta$).
\end{algorithmic}
\end{algorithm}

Theorem 1 shows that the test has size $\alpha$ under the null uniformly over $\rho$.

\begin{theorem}[Exact size for every $\rho$]\label{thm:exact}
Suppose Assumptions~\ref{ass:sutva}--\ref{ass:linear},~\ref{ass:ar1} and~\ref{ass:invariance} hold and $\mathbb{H}^{s}_0(b)$ is true. Then, conditional on potential quantities and initial conditions, for every $\alpha\in(0,1)$, $N\geq2$, $T\geq1$, $M\geq1$ and $\rho\in\R$: (i) $\Prob(p_M(\rho)\leq\alpha)\leq\alpha$; (ii) $\Prob(p^{\sup}_M\leq\alpha)\leq\alpha$; (iii) for every $\gamma\in(0,\alpha)$, $\Prob(\rho\in C_{1-\gamma})\geq1-\gamma$ and $\Prob(p^{\mathrm{BB}}_{M,\gamma}\leq\alpha)\leq\alpha$. 
\end{theorem}

Under $\mathbb{H}^{s}_0(b)$, the adjusted outcomes are fixed, so the statistic is a fixed function of the innovation array. Assumption~\ref{ass:invariance} ensures that  $(S(\varepsilon),S(g_1\varepsilon),\dots,S(g_M\varepsilon))$ exchangeable conditional on the orbit of $\varepsilon$ under $\G$, from which the results follow. Theorem~\ref{thm:exact} uses no moment or distributional assumption beyond the invariance, and no restriction on the first stage; the case of unit roots ($\rho_{0}=1$) is covered. The cost is power: $p^{\sup}_M$ exceeds $p_M(\rho_{0})$ by the amount that $p_M(\rho)$ exceeds it at wrong values of $\rho$. The confidence-set version trades $\gamma$ of level for a set that shrinks to $\rho$ as $N$ grows, as formalized in Proposition \ref{prop:power}.

\subsection{Asymptotic validity under the weak null}\label{sec-asym}

Let $N\to\infty$ with $T$ and $H$ fixed along a sequence of populations indexed by $N$; all quantities, including $\rho_{0}=\rho_{0N}\in\R$, may depend on $N$. Fix $h$ and $\theta$, write $V_{i,t+h}=V_{i,t+h}(\theta)$, $m_{t+h}=N^{-1}\sum_j\E[V_{j,t+h}]$, $\xi_i=\sum_{t\in\mathcal{T}_h}\varepsilon_{i,t}(V_{i,t+h}-m_{t+h})$, $\mu_i=\E[\xi_i]=\sum_t\Cov(\varepsilon_{i,t},V_{i,t+h})$, $v_N=N^{-1}\sum_i\Var(\xi_i)$ and $m^{2}_N=N^{-1}\sum_i\mu^{2}_i$.

\begin{assumption}[Regularity for large $N$]\label{ass:reg}
Conditional on potential quantities and initial conditions: (a) the vectors $\varepsilon_{i,1:T}$ are independent across $i$ and $\G_{\mathrm{cs}}$-invariant; (b) for some $\delta>0$ and $C<\infty$ not depending on $N$, $\E|\varepsilon_{i,t}|^{4+\delta}\leq C$, and $|Z_{i,0}|$, $|Y_{i,t}(\mathbf{0})|$, $|X_{i,t}(0)|$, $|\beta_{i,t,p}|$, $|\pi_{i,t}|$ are bounded by $C$; (c) $\liminf_Nv_N>0$, and for each $t\in\mathcal{T}_h$, $\liminf_NN^{-1}\sum_i\sigma^{2}_{i,t}>0$ and $\liminf_NN^{-1}\sum_i\E[(V_{i,t+h}-m_{t+h})^{2}]>0$.
\end{assumption}

Assumption 7 provides the regularity conditions which ensure that Lyapunov's condition hold. Because $T$ is fixed, (b) bounds the moments of $Z_{i,t}=\rho^{t}Z_{i,0}+\sum_{k\leq t}\rho^{t-k}\varepsilon_{i,k}$ uniformly over $\rho\in\R$ for any compact $\R$. I use this conditions to derive the asymptotic distribution for the p-value estimator in Theorem 2.

\begin{theorem}[Uniform asymptotic validity under the weak null]\label{thm:asym}
Suppose Assumptions~\ref{ass:sutva}--\ref{ass:linear},~\ref{ass:ar1} and~\ref{ass:reg} hold, $\G=\G_{\mathrm{cs}}$, the statistic is $S_h(r;V(\theta))$, $M=M_N\to\infty$, and $\mathbb{H}^{w}_0(h,\theta)$ is true along the sequence. Then for every sequence $\rho_N\in\R$,
\begin{equation*}
\Prob\left(p_{M}(\rho_N)\leq\alpha\right)-2\Big(1-\Phi\Big(\frac{z_{1-\alpha/2}}{\sqrt{\kappa_N}}\Big)\Big)\to0,
\qquad \kappa_N=\frac{v_N}{v_N+m^{2}_N}\in(0,1],
\end{equation*}
where $\Phi$ is the standard normal distribution function and $z_{1-\alpha/2}$ its $(1-\alpha/2)$ quantile. Consequently $\limsup_N\sup_{\rho\in\R}\Prob_\rho(p^{\sup}_M\leq\alpha)\leq\alpha$ and $\limsup_N\sup_{\rho\in\R}\Prob_\rho(p^{\mathrm{BB}}_{M,\gamma}\leq\alpha)\leq\alpha$. 
\end{theorem}

The randomization distribution of the studentized statistic converges to the standard normal, while its sampling distribution under the weak null converges to a normal with variance $\kappa_N\leq1$. The shrinkage is the usual design-based conservativeness: the unit-clustered variance estimator in~\eqref{eq:stat} estimates $\Var(\xi_i)+\mu^{2}_i$, and $\mu_i\neq0$ is possible under heterogeneity although $\sum_i\mu_i=0$ under the null \citep{jonashesh}. The uniformity over $\rho$ comes for free from the triangular-array argument, in which $\rho_N$ enters only through uniformly bounded coefficients. It is important to note that this holds only under a  fixed $T$. With $T\to\infty$ the limit law changes discontinuously at $\rho_{0}=1$ \citep{mikusheva2007} and uniform validity requires grid or hybrid constructions \citep{hansen1999,andrewsguggenberger2009}.

Validity for every $\rho$ is bought with a supremum, and the question is at what cost. Consider the sharp-null test with $\G_{\mathrm{cs}}$ and the joint statistic $S(\rho;W(b))$, and suppose the data are generated by the homogeneous linear model with true path $\beta\neq b$, so that $W_{i,t}(b)=Y_{i,t}(\mathbf{0})+\sum_p(\beta_p-b_p)X_{i,t-p}$ is stochastic. Let $\mu_h(\rho):=\lim_NN^{-1}\sum_{t\in\mathcal{T}_h}\sum_i\Cov(\varepsilon_{i,t}(\rho),W_{i,t+h}(b))$, assumed to exist; by Definition~\ref{def:filtered} each $\mu_h(\cdot)$ is an affine function of $r$, $\mu_h(\rho)=\mu_h(\rho_{0})+(\rho_{0}-\rho)\nu_h$. 

\begin{proposition}[Consistency]\label{prop:power}
Suppose Assumption~\ref{ass:reg} holds with $V$ replaced by $W(b)$ and, in (c), $\inf_{\rho\in\R}\liminf_NN^{-1}\sum_i\Var\big(\sum_t\varepsilon_{i,t}(\rho)\tilde W_{i,t+h}\big)>0$ for each $h\leq H$. Then:
(i) if the affine functions $\mu_0(\cdot),\dots,\mu_H(\cdot)$ have no common zero on $\R$, $\Prob(p^{\sup}_M\leq\alpha)\to1$ for every fixed $M\geq\alpha^{-1}-1$ and for $M=M_N\to\infty$;
(ii) under Assumption~\ref{ass:ar1} with $\liminf_NN^{-1}\sum_{i,t}\E[\tilde Z^{2}_{i,t-1}]>0$, $\sup_{\rho\in C_{1-\gamma}}|\rho-\rho_{0}|\to_p0$, so $p^{\mathrm{BB}}_{M,\gamma}-p_M(\rho_{0})\to_p\gamma$ and $\Prob(p^{\mathrm{BB}}_{M,\gamma}\leq\alpha)\to1$ whenever $\mu_h(\rho_{0})\neq0$ for some $h\leq H$.
\end{proposition}

Part (i) of Proposition \ref{prop:power} identifies the alternatives against which the maximized test is inconsistent,which are those for which some wrong persistence $\rho^\ast\in\R$ makes the adjusted outcomes orthogonal to the filtered instrument at every horizon simultaneously. For a single horizon ($H=0$) $\mu_0$ is a nonconstant affine function with a root $\rho^\ast=\rho_{0}+\mu_0(\rho_{0})/\nu_0$ whenever $\nu_0\neq0$, and the test has no power against the alternative whenever $\rho^\ast\in\R$. Part (ii) says that the confidence-set test is consistent against every alternative that the infeasible test at the true $\rho_{0}$ detects, at the cost of $\gamma$ in asymptotic level. Under local alternatives $b_N=\beta+N^{-1/2}\delta$ the vector of statistics $\{(T_h(\rho;W(b_N))/\sqrt N)_{h\leq H}:\rho\in\R\}$ converges to a Gaussian process in $\rho$ that is affine in $\rho$ with drift $\dot\mu_h(\rho)\delta$ and the asymptotic local power of the maximized test is $\Prob(\inf_{\rho\in\R}\max_hS_{h,\infty}(\rho)\geq z)$ for the limiting studentized process; this quantity is not available in closed form and we evaluate it by simulation in Section~5 rather than state it as a theorem.

\subsection{Unit effects, Stratification and Clustered Designs}\label{sec-fe}

Unit fixed effects in outcomes, treatments and instruments are pervasive in empirical applications that use short panels. The sharp-null tests handle the first two for free: under $\mathbb{H}^{s}_0(b)$ the adjusted outcomes equal $Y_{i,t}(\mathbf{0})$, which may contain arbitrary unit and period effects, and since any function of $W(b)$ is an admissible statistic, $W$ may be two-way demeaned in order to gain power. Only the innovation array is constrained by the randomization hypothesis. Unit effects in the assignment mechanism are different, because they break exchangeability across units. The next proposition shows that the uniform randomization inference procedures can be modified to remain valid under unit fixed effects.

\begin{proposition}[Unit Effects]\label{prop:fe}
Suppose Assumptions~\ref{ass:sutva}--\ref{ass:linear} hold and $\mathbb{H}^{s}_0(b)$ is true.
(i) \textbf{Unit effects in outcomes.} Theorem~\ref{thm:exact} holds with $W(b)$ replaced by any fixed transformation of it, for any $\G$ satisfying Assumption~\ref{ass:invariance}.
(ii) \textbf{Known unit means in the mechanism.} Suppose $Z_{i,t}=A_i+\rho_{0}(Z_{i,t-1}-A_i)+\varepsilon_{i,t}$ with $A_i$ observed (for instance from pre-sample data) and $\varepsilon$ satisfying Assumptions~\ref{ass:ar1} and~\ref{ass:invariance} with $\G=\G_{\mathrm{cs}}$. Then Theorems~\ref{thm:exact}--\ref{thm:asym} and Proposition~\ref{prop:filtered} hold with $Z_{i,t}$ replaced by $Z_{i,t}-A_i$.
(iii) \textbf{Unknown unit means in the mechanism.} Suppose $Z_{i,t}=A_i+\rho_{0}(Z_{i,t-1}-A_i)+\varepsilon_{i,t}$ with $A_i$ unknown and, conditional on potential quantities and initial conditions, $\varepsilon_{i,1:T}$ exchangeable within each unit. Then $\varepsilon_{i,t}(\rho_{0})=(1-\rho_{0})A_i+\varepsilon_{i,t}$ is $\G_{\mathrm{wu}}$-invariant and Theorem~\ref{thm:exact} holds with $\G=\G_{\mathrm{wu}}$ and statistics~\eqref{eq:stat} with $\tilde\varepsilon$ demeaned within units, without estimating $A_i$.
\end{proposition}

Proposition 4 is divided into three parts, each one asssociated to a different motivation for the use of unit effects. Part (i) deals with the standard inclusion of unit fixed effects because of time-invariant, individual-specific heterogeneity in the outcomes. Part (ii) is the recommended route when an external measure of the unit mean exists, as with a pre-sample climatology for a weather instrument: the mechanism becomes a known covariate-dependent map of the innovations, and nothing changes. Parts (iii) is exact and need no estimation of the unit or seasonal components: the within-unit scheme absorbs $(1-\rho_{0})A_i$ because it is constant over the permuted coordinates.

For the case in which there are unknown unit means in the assignment mechanism, if the unknown time-invariant component $A_{i}$ is fixed within known groups of units, then the procedure can be modified to perform permutations across known strata, and it remains valid.

\begin{proposition}[Stratification]\label{prop:strat}
Suppose Assumptions~\ref{ass:sutva}--\ref{ass:linear} hold and $\mathbb{H}^{s}_0(b)$ is true. Suppose that $A_{i}$ are unknown constants within known blocks $\mathcal{B}_1,\dots,\mathcal{B}_K$ of units and innovations are exchangeable across units within blocks and periods. Then,  Theorem~\ref{thm:exact} holds with the block-stratified group $\G^{\mathcal{B}}_{\mathrm{cs}}$ of within-block, within-period permutations.
\end{proposition}

It is also common to see applications in which the assignment is determined at a more aggregate level than the unit level. For instance, there might be  spatial dependence of the innovations within geographic clusters, such as microregions. If the assignment is determined at the cluster level, then the inference procedure can be modified to produce tests with exact size.

\begin{proposition}[Clustered Designs]\label{prop:cluster}
Suppose Assumptions~\ref{ass:sutva}--\ref{ass:linear} hold and $\mathbb{H}^{s}_0(b)$ is true. Suppose units are partitioned into clusters $c\in[K]$ of equal size $n$, and that, conditional on potential quantities and initial conditions, the cluster vectors $(\varepsilon_{i,t})_{i\in c}$ are exchangeable across clusters within each period, with arbitrary dependence within clusters. Then Theorem~\ref{thm:exact} holds with the cluster group $\G_{\mathrm{cl}}$ of within-period permutations of cluster labels. With unequal cluster sizes the same holds for permutations within size strata. 
\end{proposition}

Proposition \ref{prop:cluster} is divided into two cases. In the first, all clusters have the same size. The procedure remains valid as long as the permutations are restricted to observations within cluster. In the second, clusters have different sizes. In such cases, beyond doing the permutation within clusters, it is necessary to account for the different sizes of the clusters.

\section{Monte Carlo Simulations}\label{sec-mc}

In this section, I conduct Monte Carlo simulations to assess the finite-sample properties of the proposed inference procedures, and compare them with existing alternatives in the literature.

The main design is a short panel with a unit level in the instrument and a unit effect in the outcome that loads on the same level. Innovations are i.i.d.\ Gaussian with standard deviation $0.094$. The instrument is the autoregression $Z_{i,t}=\rho_{0} Z_{i,t-1}+\varepsilon_{i,t}$, so the filtered instrument at the true $\rho_{0}$ is the innovation and does not contain the unit level. The initial condition is $Z_{i,0}=A_i+e_{i,0}$, with $A_i$ of standard deviation $0.30$ and $e_{i,0}$ drawn from the innovation's stationary scale ($|\rho_{0}|<1$) or from $\mathcal{N}(0,0.094^2)$ at the unit root. The outcome is $Y_{i,t}=\beta X_{i,t}+2A_i+\alpha_i+u_{i,t}$, with $\alpha_i$ of standard deviation $0.30$ and $u$ of standard deviation $0.60$. The treatment is $X_{i,t}=\pi Z_{i,t}+v_{i,t}$ with $\pi=-18$ and $v$ of standard deviation $12$, and $\mathrm{Corr}(u,v)=0.4$. Because $\varepsilon_{i,t}(\rho)=\varepsilon_{i,t}$ is independent of $A_i$, cross-sectional permutation is exact at the true $\rho$ (Theorem~\ref{thm:exact}).

The plug-in estimates $\rho_{0}$ by the within-unit estimator and evaluates the same statistic, and a Wald test, at that point. That estimator is biased at fixed $T$ \citep{nickell1981}.

For the size exercise, we set $\beta=0$. The results are reported in Table 1.

\begin{table}[htbp]
\centering
\small
\caption{Null rejection frequencies at $\alpha=0.05$, selling design. Oracle is $p_M(\rho)$. $\hat\rho$ is the within-unit estimator. AR compares $S_0(\hat\rho)$ to the normal critical value; Wald is the plug-in $t$ test of $\theta_0=0$. $F$ is the mean unit-clustered first-stage statistic at the true $\rho$. $N=150$ except the last row, which has $N=500$. $400$ replications.}\label{tab:size}
\begin{tabular}{llccccccc}
\toprule
$T$ & $\rho_{0}$ & $\hat\rho$ & $F$ & $p_M(\rho)$ & $p^{\sup}$ & $p^{\mathrm{BB}}$ & AR & Wald \\
\midrule
$6$  & $0.5$ & $0.39$ & $19.5$ & $0.055$ & $0.000$ & $0.010$ & $0.535$ & $0.388$ \\
$6$  & $0.9$ & $0.57$ & $17.0$ & $0.060$ & $0.000$ & $0.055$ & $1.000$ & $1.000$ \\
$6$  & $1$   & $0.57$ & $15.1$ & $0.035$ & $0.013$ & $0.033$ & $1.000$ & $1.000$ \\
$12$ & $0.5$ & $0.42$ & $38.0$ & $0.045$ & $0.000$ & $0.025$ & $0.152$ & $0.102$ \\
$12$ & $0.9$ & $0.74$ & $32.9$ & $0.052$ & $0.000$ & $0.060$ & $0.995$ & $0.995$ \\
$12$ & $1$   & $0.77$ & $28.2$ & $0.045$ & $0.025$ & $0.043$ & $1.000$ & $1.000$ \\
\midrule
$6$ ($N{=}500$) & $0.9$ & $0.58$ & $52.9$ & $0.062$ & $0.048$ & $0.062$ & $1.000$ & $1.000$ \\
\bottomrule
\end{tabular}
\end{table}

The oracle p-value stays near $\alpha$, from $0.035$ to $0.062$, including the unit root case. Both the plug-in AR and the Wald procedures exhibit rejection rates greater than the desired $0.05$. When it comes to the p-values proposed in Section 4, rejection rates are closer to the desired target. The $p$-value $p^{BB}$ stays close to the target in all specifications and $p^{sup}$ lies below the target in all exercises. It is therefore valid, but much more conservative. Overall, the results of the exercise show that the $\rho$-free tests hold the null uniformly over $\rho$ and $T$, while the plug-in tests reject the null when the persistence in assignment becomes too high.

For the size exercise, I set $\beta=-0.045$. The results are summarized in Figure 1 and Table 2.

\begin{figure}[htbp]
\centering
\includegraphics[width=\textwidth]{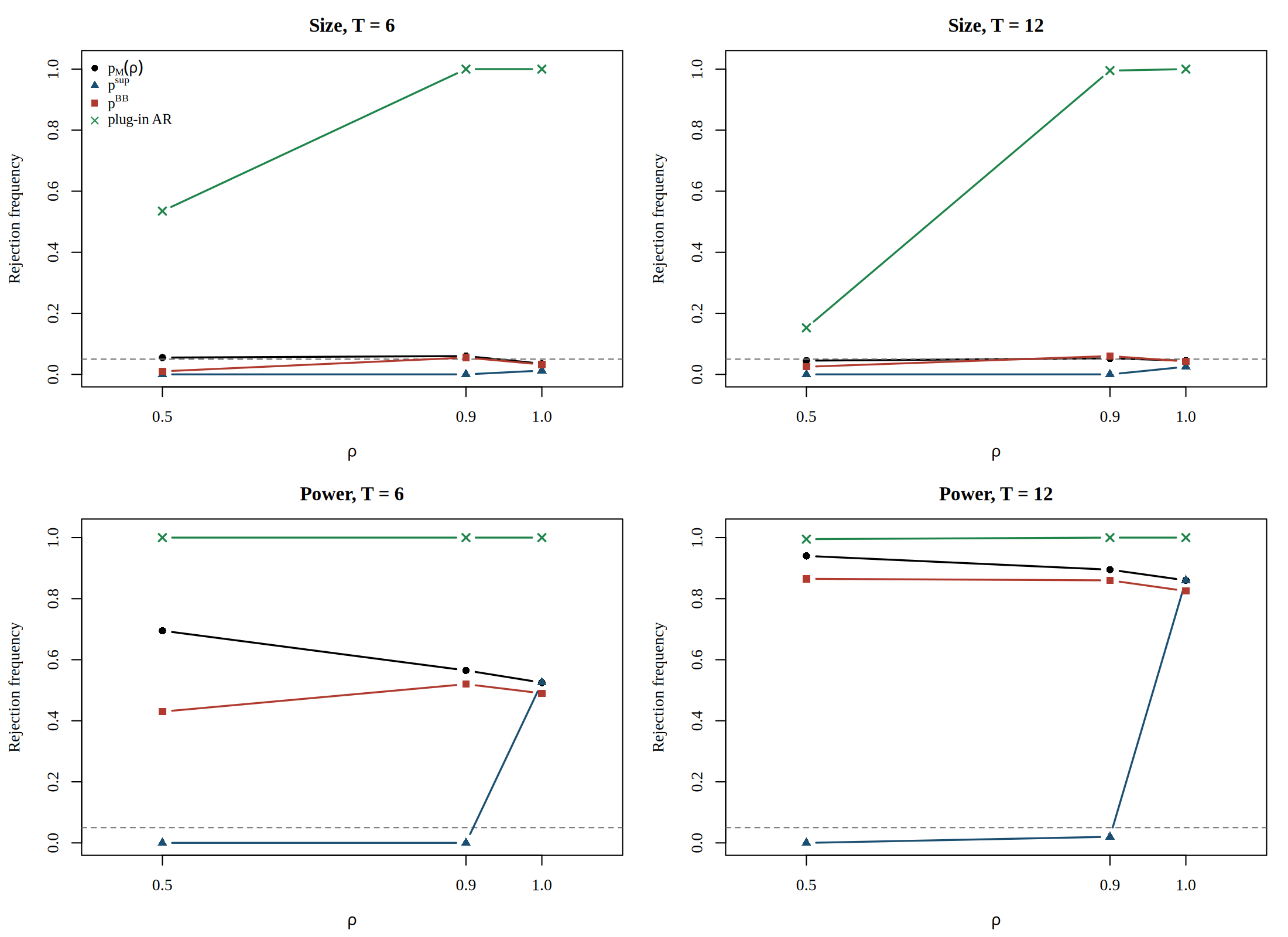}
\caption{Selling design, $N=150$. Top: null rejection frequencies. Bottom: power against $\beta=-0.045$. Black circles are $p_M(\rho)$, blue triangles $p^{\sup}$, red squares $p^{\mathrm{BB}}$, green crosses the plug-in AR test at the within estimator. The dashed line is $\alpha=0.05$.}\label{fig:mc}
\end{figure}

\begin{table}[htbp]
\centering
\small
\caption{Power at $h=0$ against $\beta=-0.045$, selling design, $200$ replications. The plug-in column is the AR test at the within estimator. $N=150$ except the last row.}\label{tab:power}
\begin{tabular}{llccccc}
\toprule
$T$ & $\rho$ & $F$ & $p_M(\rho)$ & $p^{\sup}$ & $p^{\mathrm{BB}}$ & AR \\
\midrule
$6$  & $0.5$ & $19.1$ & $0.70$ & $0.00$ & $0.43$ & $1.00$ \\
$6$  & $0.9$ & $16.8$ & $0.57$ & $0.00$ & $0.52$ & $1.00$ \\
$6$  & $1$   & $15.9$ & $0.53$ & $0.53$ & $0.49$ & $1.00$ \\
$12$ & $0.5$ & $37.3$ & $0.94$ & $0.00$ & $0.87$ & $1.00$ \\
$12$ & $0.9$ & $33.9$ & $0.90$ & $0.02$ & $0.86$ & $1.00$ \\
$12$ & $1$   & $30.2$ & $0.86$ & $0.86$ & $0.83$ & $1.00$ \\
\midrule
$6$ ($N{=}500$) & $0.9$ & $50.8$ & $0.99$ & $0.45$ & $0.98$ & $1.00$ \\
\bottomrule
\end{tabular}
\end{table}

The plug-in rejection rate of 1 in every cell is the size distortion from Table 1 carried into the alternative, so it should not be interpreted as evidence of power. The $\rho$-free tests are closest to the oracle in the exercises where the persistence is high. The supremum test, however, has no power in the cases where the persistence parameter is not the unit root.

The qualitative conclusion is that the confidence-set test delivers the oracle’s power where persistence is high, and the plain supremum does not.

\section{Empirical Application: Law Enforcement and Deforestation}\label{sec-app}

In this section, I revisit the work of \cite{agr2023} (hereafter referred to as AGR, 2023). The authors investigate the effects of a satellite law enforcement program, the DETER program, on deforestation in the Brazilian Amazon.

In 2004, the Brazilian government launched the Action Plan for the Prevention and Control of Deforestation in the Legal Amazon (PPCDAm). The program was a response to the growing concerns regarding the dangers of deforestation of tropical forests.

One of the main innovations of the PPCDAm was the development of a satellite-based system that regularly collected and processed georeferenced imagery on Amazon land cover to detect forest loss, the Real-Time System for Detection of Deforestation (DETER). Deforestation alerts that came from DETER imagery became the cornerstone for law enforcement. Officers visited alert sites and applied sanctions if evidence of illegal deforestation was found. Law enforcement officers would typically issue a fine for every environmental infraction they detected, but other forms of penalty, such as he seizure and destruction of products and equipment, are also contemplated by law.

AGR (2023) estimate the effects of law enforcement, as a function of the number of fines issued by the DETER satellite, on deforestation using panel data from 2006 to 2016. To address the issue of reverse causality between deforestation and law enforcement, the authors use an IV approach, leveraging variation in cloud coverage as exogenous variation on law enforcement. The intuition behind the strategy is that cloud coverage blocks visibility in satellite imagery and thereby limits DETER’s capacity to detect changes in land cover patterns and therefore law enforcement.

The benchmark is the annual municipality regression in \citet{agr2023}. Deforestation in year $t$, in the inverse hyperbolic sine of the PRODES increment, is regressed on the count of deforestation-related fines in year $t-1$, instrumented by the DETER cloud share in year $t-1$. The regression has municipality and year fixed effects. Precipitation and temperature are lagged one year with the instrument; PRODES cloud and non-observable area, and six municipality-weighted first-semester commodity prices, are contemporaneous. 

\begin{table}[htbp]
\centering
\small
\caption{The AGR benchmark and the filtered test at horizon $0$. Published entries are from \citet{agr2023}, Table~2. The replication is two-way fixed effects with the controls above and standard errors clustered by municipality and microregion--year. The filtered column is the paper's within-period statistic on the same lead, after municipality and year means are removed, evaluated at the Han--Phillips $\hat\rho$. $p^{\sup}$ and $p^{\mathrm{BB}}$ test $\theta_0=0$.}\label{tab:app}
\begin{tabular}{lccc}
\toprule
& Published & Replicated 2SLS & Filtered LP-IV \\
\midrule
$\theta$ on fines, IHS & $-0.050$ $(0.024)$ & $-0.050$ $(0.027)$ & $-0.036$ $(0.015)$ \\
First stage on cloud & $-9.66$ ($F=10.1$) & $-8.12$ ($F=8.1$) & \\
$\hat\rho$ (Han--Phillips) & & & $0.33$ $(0.03)$ \\
$p^{\sup}$ for $\theta_0=0$ & & & $0.35$ \\
$p^{\mathrm{BB}}$ for $\theta_0=0$ & & & $0.014$ \\
$p^{\mathrm{BB}}$ $95\%$ set & & & $[-0.069,\ -0.019]$ \\
Plug-in Wald $95\%$ & & & $[-0.065,\ -0.007]$ \\
State-mean $p^{\sup}$ / $p^{\mathrm{BB}}$ & & & $0.81$ / $0.69$ \\
\bottomrule
\end{tabular}
\end{table}

Table~\ref{tab:app} provides the comparison between the results reported in AGR, my replication of these results and the filtered 2SLS. 

The filtered estimator uses the within-period statistic on the fixed-effect residuals, so the lag of the transformed instrument is the transformed lag. $\hat\rho$ is $0.33$ (standard error $0.03$). At that value,$\hat\theta_0=-0.036$. At $\rho=0$, the same statistic returns $-0.051$, which is the unfiltered two-way IV. The randomization confidence set for $\rho$ is the grid point $\{0.35\}$. The Berger--Boos test rejects $\theta_0=0$ ($p^{\mathrm{BB}}=0.014$), and its $95\%$ set is $[-0.069,-0.019]$. The plug-in Wald interval $[-0.065,-0.007]$ also excludes zero. The supremum over the whole grid does not reject the null.

\begin{figure}[htbp]
\centering
\includegraphics[width=\textwidth]{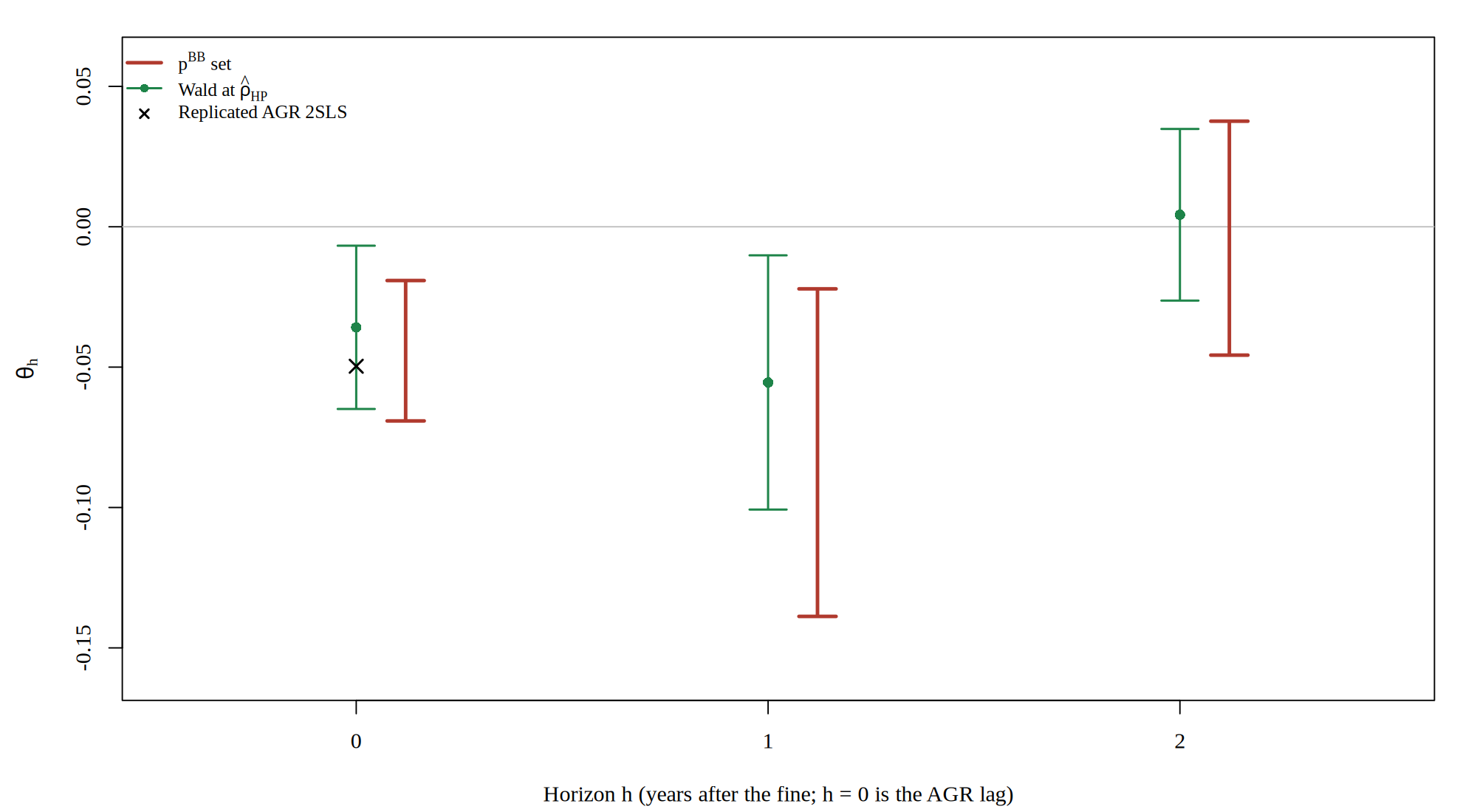}
\caption{Filtered horizon profile after municipality and year means are removed. Filled points and green intervals are the plug-in estimate and Wald interval at $\hat\rho=0.33$. Red intervals are the pointwise $95\%$ sets from $p^{\mathrm{BB}}$. The cross is the replicated AGR coefficient.}\label{fig:horizon}
\end{figure}

Figure 2 shows the dynamic effects, which can be causally interpreted under Proposition \ref{prop:filtered}. The rejection at $h=0$ agrees in sign with \citet{agr2023} and with the plug-in interval. The rejection at $h=1$ further suggests that there are dynamic effects of law enforcement in a given year, associated to fines issued up to two years before. For longer horizons, the estimates are no longer statistically significant.

\section{Conclusion}\label{sec-conclusion}

In this paper, I study inference for instrumented local projections in short panel where the instrument has an autoregressive structure. Persistence in the instrument is a nuisance in two senses: it determines which combination of dynamic causal effects the local projection estimand measures, and it governs the randomization distribution of any test built from the instrument.

I show how to conduct randomization inference that is valid for every value of the persistence parameter without estimating it, by recovering and rerandomizing the innovations of the assignment mechanism at each candidate value and reporting the worst case, and how to make the supremum computable exactly. Under a sharp null on the impulse-response path the test is exact for every persistence, including the unit root. Under a weak null on the heterogeneity-weighted estimand a studentized version is asymptotically valid uniformly over the persistence. The procedure extends to unit effects, stratified and clustered designs.

Monte Carlo simulations illustrate the desirable finite-sample properties of the proposed procedures. An empirical application shows that the law enforcement has dynamic effects on curbing deforestation of the Brazilian Amazon, but the effects fade out over time.

\bibliography{bibliography.bib}

@techreport{blandhol,
  author      = {Blandhol, Christine and Bonney, John and Mogstad, Magne and Torgovitsky, Alexander},
  title       = {When is {TSLS} Actually {LATE}?},
  institution = {National Bureau of Economic Research},
  type        = {Working Paper},
  number      = {29709},
  year        = {2022}
}

@article{bojinov,
  author  = {Bojinov, Iavor and Rambachan, Ashesh and Shephard, Neil},
  title   = {Panel Experiments and Dynamic Causal Effects: A Finite Population Perspective},
  journal = {Quantitative Economics},
  year    = {2021},
  volume  = {12},
  number  = {4},
  pages   = {1171--1196}
}

@article{ROBINS19861393,
  author  = {Robins, James},
  title   = {A New Approach to Causal Inference in Mortality Studies with a Sustained Exposure Period---Application to Control of the Healthy Worker Survivor Effect},
  journal = {Mathematical Modelling},
  year    = {1986},
  volume  = {7},
  number  = {9},
  pages   = {1393--1512}
}

@article{murphy,
  author  = {Murphy, Susan A. and van der Laan, Mark J. and Robins, James M.},
  title   = {Marginal Mean Models for Dynamic Regimes},
  journal = {Journal of the American Statistical Association},
  year    = {2001},
  volume  = {96},
  number  = {456},
  pages   = {1410--1423}
}

@article{jonashesh,
  author  = {Rambachan, Ashesh and Roth, Jonathan},
  title   = {Design-Based Uncertainty for Quasi-Experiments},
  journal = {Journal of the American Statistical Association},
  year    = {2025},
  note    = {Forthcoming, DOI 10.1080/01621459.2025.2526700}
}

@article{jorda2005,
  author  = {Jord{\`a}, {\`O}scar},
  title   = {Estimation and Inference of Impulse Responses by Local Projections},
  journal = {American Economic Review},
  year    = {2005},
  volume  = {95},
  number  = {1},
  pages   = {161--182}
}

@article{stockwatson2018,
  author  = {Stock, James H. and Watson, Mark W.},
  title   = {Identification and Estimation of Dynamic Causal Effects in Macroeconomics Using External Instruments},
  journal = {The Economic Journal},
  year    = {2018},
  volume  = {128},
  number  = {610},
  pages   = {917--948}
}

@article{plagborgwolf2021,
  author  = {Plagborg-M{\o}ller, Mikkel and Wolf, Christian K.},
  title   = {Local Projections and {VARs} Estimate the Same Impulse Responses},
  journal = {Econometrica},
  year    = {2021},
  volume  = {89},
  number  = {2},
  pages   = {955--980}
}

@article{allozagonzalosanz2025,
  author  = {Alloza, Mario and Gonzalo, Jes{\'u}s and Sanz, Carlos},
  title   = {Dynamic Effects of Persistent Shocks},
  journal = {Journal of Applied Econometrics},
  year    = {2025},
  volume  = {40},
  number  = {4},
  pages   = {380--394}
}

@unpublished{rambachanshephard,
  author = {Rambachan, Ashesh and Shephard, Neil},
  title  = {When Do Common Time Series Estimands Have Nonparametric Causal Meaning?},
  note   = {Working paper, arXiv:1903.01637},
  year   = {2021}
}

@article{herbstjohannsen2024,
  author  = {Herbst, Edward P. and Johannsen, Benjamin K.},
  title   = {Bias in Local Projections},
  journal = {Journal of Econometrics},
  year    = {2024},
  volume  = {240},
  number  = {1},
  pages   = {105655}
}

@article{meishengshi2026,
  author  = {Mei, Ziwei and Sheng, Liugang and Shi, Zhentao},
  title   = {Nickell Bias in Panel Local Projection: Financial Crises Are Worse Than You Think},
  journal = {Journal of International Economics},
  year    = {2026},
  volume  = {160},
  pages   = {104210}
}

@article{nickell1981,
  author  = {Nickell, Stephen},
  title   = {Biases in Dynamic Models with Fixed Effects},
  journal = {Econometrica},
  year    = {1981},
  volume  = {49},
  number  = {6},
  pages   = {1417--1426}
}

@article{angristkuersteiner2011,
  author  = {Angrist, Joshua D. and Kuersteiner, Guido M.},
  title   = {Causal Effects of Monetary Shocks: Semiparametric Conditional Independence Tests with a Multinomial Propensity Score},
  journal = {Review of Economics and Statistics},
  year    = {2011},
  volume  = {93},
  number  = {3},
  pages   = {725--747}
}

@article{angristjordakuersteiner2018,
  author  = {Angrist, Joshua D. and Jord{\`a}, {\`O}scar and Kuersteiner, Guido M.},
  title   = {Semiparametric Estimates of Monetary Policy Effects: String Theory Revisited},
  journal = {Journal of Business \& Economic Statistics},
  year    = {2018},
  volume  = {36},
  number  = {3},
  pages   = {371--387}
}

@article{borusyakhull2023,
  author  = {Borusyak, Kirill and Hull, Peter},
  title   = {Nonrandom Exposure to Exogenous Shocks},
  journal = {Econometrica},
  year    = {2023},
  volume  = {91},
  number  = {6},
  pages   = {2155--2185}
}

@book{fisher1935,
  author    = {Fisher, Ronald A.},
  title     = {The Design of Experiments},
  publisher = {Oliver and Boyd},
  address   = {Edinburgh},
  year      = {1935}
}

@book{lehmannromano2005,
  author    = {Lehmann, Erich L. and Romano, Joseph P.},
  title     = {Testing Statistical Hypotheses},
  edition   = {3rd},
  publisher = {Springer},
  address   = {New York},
  year      = {2005}
}

@article{andersonrubin1949,
  author  = {Anderson, Theodore W. and Rubin, Herman},
  title   = {Estimation of the Parameters of a Single Equation in a Complete System of Stochastic Equations},
  journal = {Annals of Mathematical Statistics},
  year    = {1949},
  volume  = {20},
  number  = {1},
  pages   = {46--63}
}

@article{bojinovshephard2019,
  author  = {Bojinov, Iavor and Shephard, Neil},
  title   = {Time Series Experiments and Causal Estimands: Exact Randomization Tests and Trading},
  journal = {Journal of the American Statistical Association},
  year    = {2019},
  volume  = {114},
  number  = {528},
  pages   = {1665--1682}
}

@article{shaikhtoulis2021,
  author  = {Shaikh, Azeem M. and Toulis, Panos},
  title   = {Randomization Tests in Observational Studies with Staggered Adoption of Treatment},
  journal = {Journal of the American Statistical Association},
  year    = {2021},
  volume  = {116},
  number  = {536},
  pages   = {1835--1848}
}

@article{chungromano2013,
  author  = {Chung, EunYi and Romano, Joseph P.},
  title   = {Exact and Asymptotically Robust Permutation Tests},
  journal = {Annals of Statistics},
  year    = {2013},
  volume  = {41},
  number  = {2},
  pages   = {484--507}
}

@article{diciccioromano2017,
  author  = {DiCiccio, Cyrus J. and Romano, Joseph P.},
  title   = {Robust Permutation Tests for Correlation and Regression Coefficients},
  journal = {Journal of the American Statistical Association},
  year    = {2017},
  volume  = {112},
  number  = {519},
  pages   = {1211--1220}
}

@article{wuding2021,
  author  = {Wu, Jason and Ding, Peng},
  title   = {Randomization Tests for Weak Null Hypotheses in Randomized Experiments},
  journal = {Journal of the American Statistical Association},
  year    = {2021},
  volume  = {116},
  number  = {536},
  pages   = {1898--1913}
}

@article{canayromanoshaikh2017,
  author  = {Canay, Ivan A. and Romano, Joseph P. and Shaikh, Azeem M.},
  title   = {Randomization Tests under an Approximate Symmetry Assumption},
  journal = {Econometrica},
  year    = {2017},
  volume  = {85},
  number  = {3},
  pages   = {1013--1030}
}

@article{hoeffding1951,
  author  = {Hoeffding, Wassily},
  title   = {A Combinatorial Central Limit Theorem},
  journal = {Annals of Mathematical Statistics},
  year    = {1951},
  volume  = {22},
  number  = {4},
  pages   = {558--566}
}

@article{hajek1961,
  author  = {H{\'a}jek, Jaroslav},
  title   = {Some Extensions of the {W}ald--{W}olfowitz--{N}oether Theorem},
  journal = {Annals of Mathematical Statistics},
  year    = {1961},
  volume  = {32},
  number  = {2},
  pages   = {506--523}
}

@article{neyman1923,
  author  = {Neyman, Jerzy},
  title   = {On the Application of Probability Theory to Agricultural Experiments. {E}ssay on Principles. {S}ection 9},
  journal = {Statistical Science},
  year    = {1990},
  volume  = {5},
  number  = {4},
  pages   = {465--472},
  note    = {Translated and edited by D. M. Dabrowska and T. P. Speed from the Polish original of 1923}
}

@article{dufour2006,
  author  = {Dufour, Jean-Marie},
  title   = {Monte Carlo Tests with Nuisance Parameters: A General Approach to Finite-Sample Inference and Nonstandard Asymptotics},
  journal = {Journal of Econometrics},
  year    = {2006},
  volume  = {133},
  number  = {2},
  pages   = {443--477}
}

@article{dufourkiviet1998,
  author  = {Dufour, Jean-Marie and Kiviet, Jan F.},
  title   = {Exact Inference Methods for First-Order Autoregressive Distributed Lag Models},
  journal = {Econometrica},
  year    = {1998},
  volume  = {66},
  number  = {1},
  pages   = {79--104}
}

@article{bergerboos1994,
  author  = {Berger, Roger L. and Boos, Dennis D.},
  title   = {P Values Maximized over a Confidence Set for the Nuisance Parameter},
  journal = {Journal of the American Statistical Association},
  year    = {1994},
  volume  = {89},
  number  = {427},
  pages   = {1012--1016}
}

@article{andrewsguggenberger2009,
  author  = {Andrews, Donald W. K. and Guggenberger, Patrik},
  title   = {Hybrid and Size-Corrected Subsampling Methods},
  journal = {Econometrica},
  year    = {2009},
  volume  = {77},
  number  = {3},
  pages   = {721--762}
}

@article{andrewsguggenberger2010,
  author  = {Andrews, Donald W. K. and Guggenberger, Patrik},
  title   = {Asymptotic Size and a Problem with Subsampling and with the m out of n Bootstrap},
  journal = {Econometric Theory},
  year    = {2010},
  volume  = {26},
  number  = {2},
  pages   = {426--468}
}

@article{mikusheva2007,
  author  = {Mikusheva, Anna},
  title   = {Uniform Inference in Autoregressive Models},
  journal = {Econometrica},
  year    = {2007},
  volume  = {75},
  number  = {5},
  pages   = {1411--1452}
}

@article{hansen1999,
  author  = {Hansen, Bruce E.},
  title   = {The Grid Bootstrap and the Autoregressive Model},
  journal = {Review of Economics and Statistics},
  year    = {1999},
  volume  = {81},
  number  = {4},
  pages   = {594--607}
}

@article{agr2023,
  author  = {Assun{\c{c}}{\~a}o, Juliano and Gandour, Clarissa and Rocha, Romero},
  title   = {{DETER}-ing Deforestation in the {A}mazon: Environmental Monitoring and Law Enforcement},
  journal = {American Economic Journal: Applied Economics},
  year    = {2023},
  volume  = {15},
  number  = {2},
  pages   = {125--156}
}

\section*{Appendix}

\section*{Proofs of the estimand results}

\subsection*{A covariance identity}

Under Assumption 5, for $s<t$ the variable $Z_{i,s}$ is $\mathcal{F}_{t-1}$-measurable, so $\E[\varepsilon_{i,t}Z_{i,s}]=\E[Z_{i,s}\E[\varepsilon_{i,t}\mid\mathcal{F}_{t-1}]]=0$; for $s=t$, $\E[\varepsilon_{i,t}Z_{i,t}]=\rho_{0}\E[\varepsilon_{i,t}Z_{i,t-1}]+\E[\varepsilon^{2}_{i,t}]=\sigma^{2}_{i,t}$; for $s>t$, iterating gives $Z_{i,s}=\rho_{0}^{s-t}Z_{i,t}+\sum_{k=t+1}^{s}\rho_{0}^{s-k}\varepsilon_{i,k}$ with $\E[\varepsilon_{i,t}\varepsilon_{i,k}]=\E[\varepsilon_{i,t}\E[\varepsilon_{i,k}\mid\mathcal{F}_{k-1}]]=0$ for $k>t$. Hence
\begin{equation}\label{eqS:covez}
\E[\varepsilon_{i,t}Z_{i,s}]=\rho_{0}^{\,s-t}\sigma^{2}_{i,t}\ \ (s\geq t),\qquad \E[\varepsilon_{i,t}Z_{i,s}]=0\ \ (s<t),\qquad \E[\varepsilon_{i,t}]=0.
\end{equation}

\subsection*{Proof of Proposition 1}

By Assumptions 1--3, $X_{i,t}=X_{i,t}(0)+\pi_{i,t}Z_{i,t}$ and
\begin{equation*}
Y_{i,t+h}=Y_{i,t+h}(\mathbf{0})+\sum_{p=0}^{t+h-1}\beta_{i,t+h,p}\{X_{i,t+h-p}(0)+\pi_{i,t+h-p}Z_{i,t+h-p}\}.
\end{equation*}
Since $\E[Z_{i,t}]=0$, the nonstochastic terms drop out after multiplication by $Z_{i,t}$ and expectation:
\begin{equation*}
\E[Z_{i,t}X_{i,t}]=\pi_{i,t}\Var(Z_{i,t}),\qquad
\E[Z_{i,t}Y_{i,t+h}]=\sum_{p=0}^{t+h-1}\beta_{i,t+h,p}\pi_{i,t+h-p}\Cov(Z_{i,t},Z_{i,t+h-p}).
\end{equation*}
The term $p=h$ gives $\Var(Z_{i,t})\pi_{i,t}\beta_{i,t+h,h}$. Summing over $i$ and $t\in\mathcal{T}_h$ and dividing by the denominator, which is nonzero by hypothesis, yields equation (2) of the main text. If all autocovariances vanish, the second sum is zero. $\square$

\subsection*{Proof of Proposition 2}

\emph{First stage.} $\E[\varepsilon_{i,t}X_{i,t}]=X_{i,t}(0)\E[\varepsilon_{i,t}]+\pi_{i,t}\E[\varepsilon_{i,t}Z_{i,t}]=\sigma^{2}_{i,t}\pi_{i,t}$ by \eqref{eqS:covez}; summing gives $D_h$.

\emph{Reduced form.} Multiplying the expansion of $Y_{i,t+h}$ by $\varepsilon_{i,t}$ and taking expectations, the nonstochastic terms vanish and
\begin{equation*}
\E[\varepsilon_{i,t}Y_{i,t+h}]=\sum_{p=0}^{t+h-1}\beta_{i,t+h,p}\,\pi_{i,t+h-p}\,\E[\varepsilon_{i,t}Z_{i,t+h-p}]
=\sigma^{2}_{i,t}\pi_{i,t}\beta_{i,t+h,h}+\sum_{p=0}^{h-1}\rho_{0}^{\,h-p}\sigma^{2}_{i,t}\pi_{i,t+h-p}\beta_{i,t+h,p},
\end{equation*}
because by \eqref{eqS:covez} with $s=t+h-p$ the expectation equals $\rho_{0}^{h-p}\sigma^{2}_{i,t}$ when $p\leq h$ and zero when $p>h$. Summing over $i$ and $t\in\mathcal{T}_h$ and dividing by $D_h$ gives equation (3) of the main text. Claim (i) is the vanishing of the terms $p>h$ and, for $h=0$, the absence of the second sum together with $\sum w_{i,t}=1$. Claim (ii) holds because $\sigma^{2}_{i,t}>0$, $\pi_{i,t}\geq0$ and $\rho_{0}^{h-p}\geq0$ for $\rho_{0}\geq0$. For (iii), the $w_{i,t}$ sum to one and the $\omega_{i,t,p}$ sum to the displayed expression, which vanishes exactly in the three listed cases and is strictly positive when $\rho_{0}>0$ and all $\pi_{i,t}>0$. $\square$

\subsection*{The within-unit transformed estimand}

Let $\breve Z_{i,t}=Z_{i,t}-|\mathcal{T}_h|^{-1}\sum_{s\in\mathcal{T}_h}Z_{i,s}$ and $\beta^{\mathrm{W}}_h:=\sum_{t\in\mathcal{T}_h}\sum_i\E[\breve Z_{i,t}Y_{i,t+h}]/\sum_{t\in\mathcal{T}_h}\sum_i\E[\breve Z_{i,t}X_{i,t}]$, the estimand of a local projection with unit effects in which the instrument is demeaned within units over the estimation window. Write $m_{i,t}=\E[Z_{i,t}]$ and $\breve m_{i,t}=\E[\breve Z_{i,t}]=m_{i,t}-|\mathcal{T}_h|^{-1}\sum_{s}m_{i,s}$.

\begin{proposition}[Within-unit transformed estimand]\label{propS:within}
Under Assumptions 1--3, provided the denominator is nonzero,
\begin{align*}
&\beta^{\mathrm{W}}_h\cdot\sum_{t,i}\Big\{\breve m_{i,t}X_{i,t}(0)+\pi_{i,t}\big(\Cov(\breve Z_{i,t},Z_{i,t})+\breve m_{i,t}m_{i,t}\big)\Big\}\\
&\qquad=\sum_{t,i}\breve m_{i,t}\Big\{Y_{i,t+h}(\mathbf{0})+\sum_{p=0}^{t+h-1}X_{i,t+h-p}(0)\beta_{i,t+h,p}\Big\}\\
&\qquad\quad+\sum_{t,i}\sum_{p=0}^{t+h-1}\big(\Cov(\breve Z_{i,t},Z_{i,t+h-p})+\breve m_{i,t}m_{i,t+h-p}\big)\pi_{i,t+h-p}\beta_{i,t+h,p},
\end{align*}
with sums over $t\in\mathcal{T}_h$ and $i\in[N]$. The terms involving $Y_{i,t+h}(\mathbf{0})$ and $X_{i,t+h-p}(0)$ vanish if $\breve m_{i,t}=0$ for all $(i,t)$, that is if $m_{i,t}$ is constant in $t$ within each unit, or if $Y_{i,t}(\mathbf{0})=\alpha_i+\gamma_t$ and $X_{i,t}(0)=\delta_i+\eta_t$ and the $\breve m_{i,t}$ sum to zero over $t$ for each $i$ and over $i$ for each $t$. Under Assumption 5 with $Z_{i,0}$ fixed, $m_{i,t}=\rho_{0}^{t}Z_{i,0}$, which is not constant in $t$ unless $\rho_{0}\in\{0,1\}$ or $Z_{i,0}=0$.
\end{proposition}

\begin{proof}
Expand $X_{i,t}$ and $Y_{i,t+h}$ by Assumption 3, multiply by $\breve Z_{i,t}$, take expectations, and use $\E[\breve Z_{i,t}U]=\Cov(\breve Z_{i,t},U)+\breve m_{i,t}\E[U]$ for any $U$ and $\E[\breve Z_{i,t}]c=\breve m_{i,t}c$ for nonstochastic $c$. The vanishing conditions are immediate; $\sum_{t\in\mathcal{T}_h}\breve m_{i,t}=0$ always holds by construction, so the additive case needs only the within-period sum condition for the period effects, which holds if $\breve m_{i,t}$ does not depend on $i$. Under Assumption 5, $\E[Z_{i,t}\mid Z_{i,0}]=\rho_{0}\E[Z_{i,t-1}\mid Z_{i,0}]$, hence $m_{i,t}=\rho_{0}^{t}Z_{i,0}$.
\end{proof}

The two-way transformed estimand, which also demeans within periods, is treated identically with $\breve Z$ replaced by the two-way transformed instrument; the within-period demeaning removes components of $m_{i,t}$ common to all units but not the unit-specific component $\rho_{0}^{t}Z_{i,0}$.

\section*{Proofs of the inference results}

\subsection*{Proof of Theorem 1}

Under $\mathbb{H}^{s}_0(b)$ the array $W=W(b)$ equals $(Y_{i,t}(\mathbf{0}))$ and is fixed, so at the true $\rho_{0}$ the statistic is a fixed function $\varepsilon\mapsto S(\varepsilon;W)$ of the innovation array $\varepsilon=\varepsilon(\rho_{0})$.

\emph{(i)} Let $\mathcal{O}(\varepsilon)=\{g\varepsilon:g\in\G\}$ be the orbit of $\varepsilon$, as a multiset of $|\G|$ elements. For any $g\in\G$, $\mathcal{O}(g\varepsilon)=\mathcal{O}(\varepsilon)$, so by Assumption 6, $(g\varepsilon,\mathcal{O}(\varepsilon))\overset{d}{=}(\varepsilon,\mathcal{O}(\varepsilon))$ and hence, conditional on the orbit, $\varepsilon$ is uniformly distributed on the multiset $\mathcal{O}(\varepsilon)$ \citep[Section 15.2]{lehmannromano2005}. The draws $g_1,\dots,g_M$ are i.i.d. uniform on $\G$ and independent of $\varepsilon$; conditional on $(\mathcal{O}(\varepsilon),\varepsilon)$ each $g_m\varepsilon$ is uniform on $\mathcal{O}(\varepsilon)$, independently across $m$, with a conditional law not depending on $\varepsilon$. Hence conditional on the orbit, $(\varepsilon,g_1\varepsilon,\dots,g_M\varepsilon)$ are i.i.d. uniform on $\mathcal{O}(\varepsilon)$ and $(S_0,\dots,S_M):=(S(\varepsilon;W),S(g_1\varepsilon;W),\dots,S(g_M\varepsilon;W))$ is exchangeable. Let $R$ be the number of coordinates at least as large as $S_0$, including $S_0$; then $p_M(\rho_{0})=R/(M+1)$. For an exchangeable vector, $\Prob(R\leq k)\leq k/(M+1)$ for $k=1,\dots,M+1$ (equality without ties; ties increase $R$). Therefore $\Prob(p_M(\rho_{0})\leq\alpha)=\Prob(R\leq\lfloor\alpha(M+1)\rfloor)\leq\alpha$, conditionally on the orbit and hence unconditionally.

\emph{(ii)} Since $\rho_{0}\in\R$ and the draws are common to all $\rho$, $p^{\sup}_M\geq p_M(\rho_{0})$ realization by realization, so $\{p^{\sup}_M\leq\alpha\}\subseteq\{p_M(\rho_{0})\leq\alpha\}$.

\emph{(iii)} At $\rho=\rho_{0}$ the auxiliary statistic $B$ of equation (7), computed with the lagged instrument rebuilt from the innovations and $Z_{1:N,0}$, is a fixed function of $\varepsilon$, so (i) applied to $B$ gives $\Prob(p^{\rho_{0}}_M(\rho_{0})\leq\gamma)\leq\gamma$, i.e. $\Prob(\rho_{0}\in C_{1-\gamma})\geq1-\gamma$. Then
\begin{align*}
\Prob(p^{\mathrm{BB}}_{M,\gamma}\leq\alpha)&\leq\Prob\Big(\sup_{r\in C_{1-\gamma}}p_M(\rho)\leq\alpha-\gamma,\ \rho_{0}\in C_{1-\gamma}\Big)+\Prob(\rho_{0}\notin C_{1-\gamma})\\
&\leq\Prob(p_M(\rho_{0})\leq\alpha-\gamma)+\gamma\leq\alpha.
\end{align*}
The uniformity and coverage statements follow since the bounds hold for each $\rho_{0}\in\R$. $\square$

\subsection*{Proof of Theorem 2}

All limits are as $N\to\infty$ along an arbitrary sequence of populations satisfying Assumption 8 with $\rho_{0N}\in\R$; the index $N$ on $\rho_{0}$ and population quantities is suppressed, and $C$ denotes a constant not depending on $N$ that may change from line to line. Write $\varepsilon_{i,t}=\varepsilon_{i,t}(\rho_{0})$, $V_{i,t+h}=V_{i,t+h}(\theta)$, $a_{i,t}=\tilde\varepsilon_{i,t}$, $c_{i,t}=\tilde V_{i,t+h}$, $T_N=\sum_{t\in\mathcal{T}_h}\sum_ia_{i,t}c_{i,t}$ and $\hat s^{2}_N=\sum_i(\sum_ta_{i,t}c_{i,t})^{2}$.

\emph{Step 0 (moment bounds).} Since $T$ is fixed and $\R$ compact, $Z_{i,t}=\rho_{0}^{t}Z_{i,0}+\sum_{k=1}^{t}\rho_{0}^{t-k}\varepsilon_{i,k}$ satisfies $\E|Z_{i,t}|^{4+\delta}\leq C$ uniformly in $i,t,N$ and $\rho_{0}\in\R$ by Assumption 8(b) and the $c_r$-inequality. $V_{i,t+h}$ is affine in $(Z_{i,1},\dots,Z_{i,t+h})$ with coefficients bounded by $C$, so $\E|V_{i,t+h}|^{4+\delta}\leq C$ and $\E|\varepsilon_{i,t}V_{i,s}|^{2+\delta/2}\leq C$. The vectors $(\varepsilon_{i,1:T},V_{i,1:T})$ are independent across $i$.

\emph{Step 1 (linearization).} With $m_{t+h}=N^{-1}\sum_j\E[V_{j,t+h}]$ and $\xi_i=\sum_t\varepsilon_{i,t}(V_{i,t+h}-m_{t+h})$, since $\sum_ia_{i,t}c_{i,t}=\sum_i\varepsilon_{i,t}(V_{i,t+h}-\bar V_{t+h})$,
$T_N-\sum_i\xi_i=\sum_{t\in\mathcal{T}_h}(\sum_i\varepsilon_{i,t})(m_{t+h}-\bar V_{t+h})$. By independence and $\E[\varepsilon_{i,t}]=0$, $\sum_i\varepsilon_{i,t}=O_p(N^{1/2})$, and $\bar V_{t+h}-m_{t+h}=O_p(N^{-1/2})$. Hence $T_N=\sum_i\xi_i+O_p(1)$.

\emph{Step 2 (the null).} $\mu_i=\E[\xi_i]=\sum_t\Cov(\varepsilon_{i,t},V_{i,t+h})=\sum_t\{\E[\varepsilon_{i,t}Y_{i,t+h}]-\theta\E[\varepsilon_{i,t}X_{i,t}]\}$, so by the proof of Proposition 2, $\sum_i\mu_i=D_h(\theta_h-\theta)=0$ under $\mathbb{H}^{w}_0(h,\theta)$.

\emph{Step 3 (CLT).} The $\xi_i$ are independent with $\E|\xi_i-\mu_i|^{2+\delta/2}\leq C$, and $v_N\in[c,C]$ by Assumption 8(c). Lyapunov's condition holds, so $(Nv_N)^{-1/2}\sum_i(\xi_i-\mu_i)\Rightarrow N(0,1)$ and, by Steps 1--2, $(Nv_N)^{-1/2}T_N\Rightarrow N(0,1)$.

\emph{Step 4 (variance estimator).} Write $\sum_ta_{i,t}c_{i,t}=\xi_i+R_i$ with $R_i=\sum_t\{-\bar\varepsilon_t(V_{i,t+h}-m_{t+h})-\varepsilon_{i,t}(\bar V_{t+h}-m_{t+h})+\bar\varepsilon_t(\bar V_{t+h}-m_{t+h})\}$. Since $\bar\varepsilon_t$ and $\bar V_{t+h}-m_{t+h}$ are $O_p(N^{-1/2})$ while $N^{-1}\sum_i(V_{i,t+h}-m_{t+h})^{2}$ and $N^{-1}\sum_i\varepsilon^{2}_{i,t}$ are $O_p(1)$, $N^{-1}\sum_iR^{2}_i=O_p(N^{-1})$ and $N^{-1}\sum_i\xi_iR_i=O_p(N^{-1/2})$ by Cauchy--Schwarz. The $\xi^{2}_i$ are independent with $\E|\xi^{2}_i|^{1+\delta/4}\leq C$, so by the weak law of large numbers for independent arrays with uniformly bounded $(1+\delta')$ moments (e.g. via the von Bahr--Esseen inequality), $N^{-1}\sum_i(\xi^{2}_i-\E\xi^{2}_i)\to_p0$, and $\E\xi^{2}_i=\Var(\xi_i)+\mu^{2}_i$. Hence $N^{-1}\hat s^{2}_N=v_N+m^{2}_N+o_p(1)$ with $m^{2}_N\leq C$, and $S_N:=|T_N|/\hat s_N=\sqrt{\kappa_N}\,|(Nv_N)^{-1/2}T_N|+o_p(1)$ with $\kappa_N=v_N/(v_N+m^{2}_N)\in[c',1]$. By uniform continuity of $\Phi$ and compactness of $[c',1]$, $\Prob(S_N\geq x)-2(1-\Phi(x/\sqrt{\kappa_N}))\to0$ for every $x>0$.

\emph{Step 5 (randomization distribution).} Condition on the data $\mathcal{D}$ and let $g=(\sigma_1,\dots,\sigma_T)$ be uniform on $\G_{\mathrm{cs}}$; $\E^{*}$, $\Var^{*}$, $\Prob^{*}$ denote conditional moments and probabilities. The permuted statistic is $T^{*}_N=\sum_{t\in\mathcal{T}_h}L_t$, $L_t=\sum_ia_{\sigma_t(i),t}c_{i,t}$, with the $L_t$ conditionally independent across $t$. Since $\sum_ia_{i,t}=\sum_ic_{i,t}=0$, $\E^{*}[L_t]=0$ and $\Var^{*}(L_t)=(N-1)^{-1}\sum_ia^{2}_{i,t}\sum_ic^{2}_{i,t}$ \citep{hoeffding1951}. By Step 0 and the weak law of large numbers, $N^{-1}\sum_ia^{2}_{i,t}-N^{-1}\sum_i\sigma^{2}_{i,t}\to_p0$ and $N^{-1}\sum_ic^{2}_{i,t}-N^{-1}\sum_i\E(V_{i,t+h}-m_{t+h})^{2}\to_p0$, both limits bounded away from zero by Assumption 8(c), and the standardized fourth moments $N^{-1}\sum_ia^{4}_{i,t}/(N^{-1}\sum_ia^{2}_{i,t})^{2}$ and $N^{-1}\sum_ic^{4}_{i,t}/(N^{-1}\sum_ic^{2}_{i,t})^{2}$ are $O_p(1)$. Hence the Lyapunov-type condition
\begin{equation}\label{eqS:lyap}
\frac{1}{N}\cdot\frac{N^{-1}\sum_ia^{4}_{i,t}}{(N^{-1}\sum_ia^{2}_{i,t})^{2}}\cdot\frac{N^{-1}\sum_ic^{4}_{i,t}}{(N^{-1}\sum_ic^{2}_{i,t})^{2}}\to_p0
\end{equation}
holds, and it implies the Lindeberg-type condition of \citet{hajek1961} for $L_t$ (apply Markov's inequality to the indicator in the Lindeberg sum). By the combinatorial central limit theorem of \citet{hoeffding1951} in the form of \citet{hajek1961}, $L_t/\sqrt{\Var^{*}(L_t)}$ converges in conditional law to $N(0,1)$ in probability for each $t$, and since the $L_t$ are conditionally independent and finitely many with $v^{*}_N:=N^{-1}\Var^{*}(T^{*}_N)$ bounded away from zero in probability, $\sup_x|\Prob^{*}(T^{*}_N/\sqrt{Nv^{*}_N}\leq x)-\Phi(x)|\to_p0$. For $\hat s^{*2}_N=\sum_i(\sum_ta_{\sigma_t(i),t}c_{i,t})^{2}$, expand the square: the diagonal terms $N^{-1}\sum_ia^{2}_{\sigma_t(i),t}c^{2}_{i,t}$ have conditional mean $(N^{-1}\sum_ja^{2}_{j,t})(N^{-1}\sum_ic^{2}_{i,t})$ and conditional variance at most $(N-1)^{-1}(N^{-1}\sum_ia^{4}_{i,t})(N^{-1}\sum_ic^{4}_{i,t})=O_p(N^{-1})$ by the permutation variance formula; the cross terms $N^{-1}\sum_ia_{\sigma_t(i),t}a_{\sigma_s(i),s}c_{i,t}c_{i,s}$, $t\neq s$, have conditional mean zero by independence of $\sigma_t$ and $\sigma_s$ and $\sum_ia_{i,t}=0$, and conditional second moment $O_p(N^{-1})$ (the $i=j$ terms contribute $N^{-2}\sum_ic^{2}_{i,t}c^{2}_{i,s}\E^{*}[a^{2}_{\sigma_t(i),t}]\E^{*}[a^{2}_{\sigma_s(i),s}]=O_p(N^{-1})$ and the $i\neq j$ terms involve $\E^{*}[a_{\sigma_t(i),t}a_{\sigma_t(j),t}]=-(N(N-1))^{-1}\sum_ka^{2}_{k,t}$ twice, contributing $O_p(N^{-2})$). Hence $N^{-1}\hat s^{*2}_N-v^{*}_N\to_p0$ and
$\sup_x|\Prob^{*}(S^{*}_N\leq x)-(2\Phi(x)-1)|\to_p0$ for $S^{*}_N=|T^{*}_N|/\hat s^{*}_N$.

\emph{Step 6 (the $p$-value).} Let $\hat F^{*}_N(x)=\Prob^{*}(S^{*}_N\geq x)$ and $\hat F^{*}_{N,M}(x)=M^{-1}\sum_m\ind\{S^{(m)}\geq x\}$ its Monte Carlo version from $M=M_N\to\infty$ i.i.d. draws; by the Dvoretzky--Kiefer--Wolfowitz inequality $\sup_x|\hat F^{*}_{N,M}(x)-\hat F^{*}_N(x)|\to_p0$. Since $p_M(\rho_{0})=(1+M\hat F^{*}_{N,M}(S_N))/(M+1)$, Step 5 gives $p_M(\rho_{0})=2(1-\Phi(S_N))+o_p(1)$, and $\{p_M(\rho_{0})\leq\alpha\}$ differs from $\{S_N\geq z_{1-\alpha/2}\}$ by an event of vanishing probability because the limit law of $S_N$ is continuous. By Step 4, $\Prob(p_M(\rho_{0})\leq\alpha)-2(1-\Phi(z_{1-\alpha/2}/\sqrt{\kappa_N}))\to0$; the limit is at most $\alpha$ since $\kappa_N\leq1$, with equality along the sequence iff $\kappa_N\to1$, i.e. $m_N\to0$. Under $\beta_{i,t,p}=\beta_p$ and $\pi_{i,t}=\pi$, the proof of Corollary 2 gives $\mu_i=\pi\sum_t\sigma^{2}_{i,t}(\sum_{p\leq h}\rho_{0}^{h-p}\beta_p-\theta)=\pi\sum_t\sigma^{2}_{i,t}(\theta_h-\theta)=0$ under the null, so $m_N=0$.

\emph{Step 7 (uniformity; $\rho_{}$-free $p$-values).} Step 6 holds for every sequence of populations and every $\rho_{N}\in\R$ satisfying Assumption 8. If $\limsup_N\sup_{\rho_{0}\in\R}\Prob_{\rho_{0}}(p^{\sup}_M\leq\alpha)>\alpha$, there are $\eta>0$, a subsequence $N_k$ and $\rho_{0N_k}\in\R$ with $\Prob_{\rho_{0N_k}}(p^{\sup}_M\leq\alpha)\geq\alpha+\eta$; but $p^{\sup}_M\geq p_M(\rho_{0N_k})$ realization by realization, so the left side is at most $\Prob_{\rho_{0N_k}}(p_M(\rho_{0N_k})\leq\alpha)$, whose limit along the subsequence is at most $\alpha$ by Step 6, a contradiction. For $p^{\mathrm{BB}}_{M,\gamma}$, Theorem 1(iii) gives $\Prob_{\rho_{0}}(\rho_{0}\notin C_{1-\gamma})\leq\gamma$ exactly, since the auxiliary test concerns the innovation array only, so $\Prob_{\rho_{0}}(p^{\mathrm{BB}}_{M,\gamma}\leq\alpha)\leq\Prob_{\rho_{0}}(p_M(\rho_{0})\leq\alpha-\gamma)+\gamma$ and the same argument at level $\alpha-\gamma$ gives the bound $\alpha$. $\square$

\begin{remark}
The proof relies on two cited results: the combinatorial central limit theorem for linear permutation statistics under the condition \eqref{eqS:lyap} \citep{hoeffding1951,hajek1961}, and a weak law of large numbers for independent arrays with uniformly bounded $(1+\delta')$ moments. With $M$ fixed the same argument gives asymptotic level $\lfloor\alpha(M+1)\rfloor/(M+1)$. The joint statistic $\max_hS_h$ under a joint weak null on $(\theta_0,\dots,\theta_H)$ requires a multivariate version of Step 5, which is routine.
\end{remark}

\subsection*{Proof of Proposition 3}

Throughout, the data are generated by the homogeneous linear model with path $\beta\neq b$, so $W_{i,t}(b)=Y_{i,t}(\mathbf{0})+\sum_p(\beta_p-b_p)X_{i,t-p}$ is affine in $(Z_{i,1},\dots,Z_{i,t})$ with bounded coefficients, and the moment bounds of Step 0 of the proof of Theorem 3 apply to $W(b)$ in place of $V$. Write $\bar\rho_{0}=\sup\{|\rho|:\rho\in\R\}$.

\emph{(i) The observed statistics diverge uniformly in $\rho$.} By the proof of Lemma 1, $T_h(\rho;W)=A_h-\rho B_h$ with $A_h=\sum_{t,i}\tilde Z_{i,t}\tilde W_{i,t+h}$ and $B_h=\sum_{t,i}\tilde Z_{i,t-1}\tilde W_{i,t+h}$. By the linearization of Step 1 applied to the pairs $(Z_{\cdot,t},W_{\cdot,t+h})$ and $(Z_{\cdot,t-1},W_{\cdot,t+h})$, $A_h=\E[A_h]+O_p(N^{1/2})$ and $B_h=\E[B_h]+O_p(N^{1/2})$, where $N^{-1}\E[A_h]-N^{-1}\E[B_h]\,\rho\to\mu_h(\rho)$ uniformly on $\R$ (the convergence is of two scalar sequences). Hence $\sup_{\rho\in\R}|T_h(\rho;W)-N\mu_h(\rho)|=O_p(N^{1/2})$. Also $\hat s^{2}_h(\rho;W)=\sum_i(a_{h,i}-rb_{h,i})^{2}\leq2\sum_i(a^{2}_{h,i}+\bar\rho_{0}^{2}b^{2}_{h,i})=O_p(N)$ uniformly in $\rho$. If $\mu_0,\dots,\mu_H$ have no common zero on $\R$, then $\underline m:=\inf_{\rho\in\R}\max_h|\mu_h(\rho)|>0$ by continuity and compactness, and therefore $\inf_{\rho\in\R}S(\rho;W)\geq\underline m N^{1/2}/C-O_p(1)\to_p\infty$.

\emph{The permuted statistics are bounded uniformly in $\rho$.} For a draw $g$, $S^{g}_h(\rho)=|A_{g,h}-\rho B_{g,h}|/\hat s^{g}_h(\rho)$. Conditional on the data, $A_{g,h}$ and $B_{g,h}$ are sums over $t$ of linear permutation statistics with conditional mean zero and conditional variance $(N-1)^{-1}\sum_i\tilde Z^{2}_{i,t}\sum_i\tilde W^{2}_{i,t+h}$ and similarly with $\tilde Z_{i,t-1}$; by Step 0 these are $O_p(N)$, so $\E^{*}[\sup_\rho(A_{g,h}-\rho B_{g,h})^{2}]\leq2\E^{*}[A^{2}_{g,h}]+2\bar\rho_{0}^{2}\E^{*}[B^{2}_{g,h}]=O_p(N)$. For the denominator, $N^{-1}\hat s^{2,g}_h(\rho)=N^{-1}\sum_i(a_{g,h,i}-\rho b_{g,h,i})^{2}$ is a quadratic polynomial in $\rho$ whose three coefficients are of the form $N^{-1}\sum_iu_{g,i}v_{g,i}$ with $u,v\in\{a,b\}$; by the argument of Step 5 of the proof of Theorem 3 each coefficient concentrates around its conditional mean at rate $O_p(N^{-1/2})$, and the conditional mean of $N^{-1}\hat s^{2,g}_h(\rho)$ equals $\sum_t(N-1)^{-1}\sum_j\tilde\varepsilon^{2}_{j,t}(\rho)\sum_i\tilde W^{2}_{i,t+h}/N+o_p(1)$, which is bounded below uniformly on $\R$ in probability by the nondegeneracy hypothesis of the proposition. Hence $\inf_{\rho\in\R}N^{-1}\hat s^{2,g}_h(\rho)\geq c>0$ with conditional probability tending to one, and $\E^{*}[\sup_{\rho\in\R}S^{g}(\rho)^{2}\,\ind\{\inf_\rho N^{-1}\hat s^{2,g}\geq c\}]=O_p(1)$.

\emph{Conclusion of (i).} Let $x_N=\inf_{\rho\in\R}S(\rho;W)\to_p\infty$. For fixed $M$, $\Prob(\max_{m\leq M}\sup_\rho S^{g_m}(\rho)<x_N)\to1$, on which event $p_M(\rho)=1/(M+1)\leq\alpha$ for all $\rho\in\R$, so $p^{\sup}_M\leq\alpha$. For $M\to\infty$, the fraction of draws with $\sup_\rho S^{g_m}(\rho)\geq x_N$ has conditional mean $\Prob^{*}(\sup_\rho S^{g}(\rho)\geq x_N)\leq O_p(1)/x^{2}_N+o_p(1)\to_p0$ by Markov's inequality, so $p^{\sup}_M\leq(1+Mo_p(1))/(M+1)\to_p0$.

\emph{(ii) The confidence set shrinks.} By equation (7), $B(\rho)=|\sum_{t,i}\tilde\varepsilon_{i,t}\tilde Z_{i,t-1}+(\rho_{0}-\rho)\sum_{t,i}\tilde Z^{2}_{i,t-1}|$, where the first sum is $O_p(N^{1/2})$ by Step 1 (its mean is zero by \eqref{eqS:covez}) and the second equals $N\kappa_N+O_p(N^{1/2})$ with $\liminf\kappa_N>0$ by hypothesis. Hence for any $\eta>0$, $\inf_{|\rho-\rho_{0}|\geq\eta,\,r\in\R}B(\rho)\geq\eta N\kappa_N/2$ with probability tending to one. The transformed statistic $B(g\varepsilon(\rho))=|\sum_{t,i}\widetilde{(g\varepsilon(\rho))}_{i,t}\tilde Z^{*}_{i,t-1}(\rho)|$ uses the path rebuilt from the permuted innovations, $Z^{*}_{i,t-1}(\rho)=\rho^{t-1}Z_{i,0}+\sum_{k<t}r^{t-1-k}(g\varepsilon(\rho))_{i,k}$, a polynomial in $\rho$ with coefficients that are permuted innovations; by Step 0 and $T$ fixed, $\E^{*}[\sup_{\rho\in\R}B(g\varepsilon(\rho))^{2}]=O_p(N)$. By Markov's inequality the fraction of draws with $B(g_m\varepsilon(\rho))\geq\eta N\kappa_N/2$ is $o_p(1)$ uniformly over $|\rho-\rho_{0}|\geq\eta$, so $p^{\rho_{0}}_M(\rho)\to_p1/(M+1)\leq\gamma$ uniformly on that set (for $M\geq\gamma^{-1}-1$), and $C_{1-\gamma}\subseteq(\rho_{0}-\eta,\rho_{0}+\eta)$ with probability tending to one. Since $\eta$ is arbitrary, $\sup_{\rho\in C_{1-\gamma}}|\rho-\rho_{0}|\to_p0$. For $\rho$ with $|\rho-\rho_{0}|\leq\eta$, $T_h(\rho;W)-T_h(\rho_{0};W)=(\rho_{0}-\rho)B_h$ with $B_h=O_p(N)$ under the alternative and $O_p(N^{1/2})$ under the null, and $\hat s_h(\rho)-\hat s_h(\rho_{0})=O_p(\eta N^{1/2})$; hence under the null $\sup_{|\rho-\rho_{0}|\leq\eta_N}|S_h(\rho)-S_h(\rho_{0})|\to_p0$ for any $\eta_N\to0$, which with the continuity of the limiting randomization distribution (Step 5) gives $p^{\mathrm{BB}}_{M,\gamma}-p_M(\rho_{0})\to_p\gamma$; and under an alternative with $\mu_h(\rho_{0})\neq0$ for some $h$, choosing $\eta<|\mu_h(\rho_{0})|/(2|\nu_h|)$ (any $\eta$ if $\nu_h=0$) gives $\inf_{|\rho-\rho_{0}|\leq\eta}S_h(\rho;W)\to_p\infty$ by part (i), so $\Prob(p^{\mathrm{BB}}_{M,\gamma}\leq\alpha)\to1$. $\square$

\subsection*{Proof of Proposition 4}

\emph{(i)} In the proof of Theorem 1 the statistic was an arbitrary fixed function of the innovation array; $W(b)$ enters only as a fixed array, so any fixed transformation of it, including two-way demeaning, is admissible.

\emph{(ii)} Define $Z'_{i,t}=Z_{i,t}-A_i$. Then $Z'_{i,t}=\rho_{0} Z'_{i,t-1}+\varepsilon_{i,t}$ with the same innovations, so Assumption 5 holds for $Z'$, and the filtered instrument at $r$ computed from $Z'$ is $\varepsilon_{i,t}(\rho)=(Z_{i,t}-A_i)-\rho(Z_{i,t-1}-A_i)$. Since $A_i$ is known, $Z'$ is observed, and all results apply to the panel $(Z',X,Y)$ verbatim. For Proposition 2, note that $X_{i,t}=X_{i,t}(0)+\pi_{i,t}Z_{i,t}=\{X_{i,t}(0)+\pi_{i,t}A_i\}+\pi_{i,t}Z'_{i,t}$, so Assumption 3 holds for $Z'$ with modified untreated treatments, and the decomposition (3) is unchanged.

\emph{(iii)} At $\rho=\rho_{0}$, $\varepsilon_{i,t}(\rho_{0})=Z_{i,t}-\rho_{0} Z_{i,t-1}=(1-\rho_{0})A_i+\varepsilon_{i,t}$. For $g=(\tau_1,\dots,\tau_N)\in\G_{\mathrm{wu}}$, $(g\varepsilon(\rho_{0}))_{i,t}=(1-\rho_{0})A_i+\varepsilon_{i,\tau_i(t)}$, which has the same conditional law as $\varepsilon(\rho_{0})$ by the within-unit exchangeability of $\varepsilon_{i,1:T}$, since the shift $(1-\rho_{0})A_i$ is constant across the permuted coordinates. Thus Assumption 6 holds for the array $\varepsilon(\rho_{0})$ with $\G=\G_{\mathrm{wu}}$, and the proof of Theorem 1 applies without change (it uses only the invariance of the array that is permuted and the fixedness of $W$).

\subsection*{Proof of Proposition 5}

The proof is similar to the proof of Proposition 4 (iii). For the block version, $(g\varepsilon(\rho_{0}))_{i,t}=(1-\rho_{0})A_{\sigma_t(i)}+\varepsilon_{\sigma_t(i),t}$ with $\sigma_t$ a within-block permutation; since $A$ is constant within blocks, $A_{\sigma_t(i)}=A_i$, and within-block exchangeability of the innovations gives the invariance.

\subsection*{Proof of Proposition 6}

Fix an ordering of units within each cluster and let $\G_{\mathrm{cl}}$ act by $(g\varepsilon)_{(c,j),t}=\varepsilon_{(\sigma_t(c),j),t}$ for $j\leq n$, where $\sigma_t$ permutes cluster labels. Exchangeability of the ordered cluster vectors across clusters within periods is Assumption 6 for this group (an example is $\varepsilon_{i,t}=u_{c(i),t}+e_{i,t}$ with $(u_{c,t})$ i.i.d. across $c$ and $(e_{i,t})$ i.i.d. across $i$, for which any ordering gives the same joint law). Theorem 1 then applies. With unequal sizes, permutations within size strata are a subgroup and the same argument applies. For the weak null, the proof of Theorem 3 goes through with clusters as the independent units: Step 3 is a Lyapunov CLT over $K\to\infty$ independent cluster sums $\xi_c=\sum_{i\in c}\xi_i$, Step 4 uses the cluster-clustered variance $\sum_c(\sum_{i\in c}\sum_ta_{i,t}c_{i,t})^{2}$, and Step 5 is the combinatorial CLT for the linear permutation statistic $\sum_c\langle a_{\sigma_t(c),t},c_{c,t}\rangle$ with $n$-dimensional ``scores'', to which \citet{hoeffding1951} and \citet{hajek1961} apply coordinatewise by linearity, since $n$ is fixed.
\end{document}